# THE UNITARY CLOTHING TRANSFORMATION METHOD IN QUANTUM FIELD THEORY. RENORMALIZATION OF COUPLING CONSTANT

MASTER THESIS

**Yeletskikh Ivan Vladimirovich**

6 course graduate student

Department of Theoretical Nuclear Physics

Scientific advisor:

D.Sc., Head of Department

in Institute of Electrophysics

and Radiation Thechnologies

NAS of Ukraine

............................................................................................... V.Yu.Korda

Kharkov 2007

Аннотация

С помощью метода унитарных одевающих преобразований в модели квантовой теории поля, включающей заряженное бесспиновое нуклонное и нейтральное мезонное поля, связанные трилинейным взаимодействием типа Юкавы, получено выражение для поправки к величине заряда в третьем порядке по константе связи. Будучи определенным вне энергетической оболочки, найденное выражение может быть представлено на энергетической оболочке в явно ковариантной форме, что обеспечивает независимость от импульсов частиц. Сравнение с соответствующим результатом ковариантной теории возмущений Дайсона осуществлено с помощью установленной связи между старомодной теорией возмущений и методом одевающих преобразований.

Анотація

За допомогою методу унітарних одягаючих перетворень в моделі зарядженого безспінового нуклонного і нейтрального мезонного полів, які зв'язані трилінійною взаємодією типу Юкави, знайдено вираз для поправки до величини заряду в третьому порядку за константою зв'язку. Через те що розрахований вираз визначено поза енергетичною оболонкою, його можна подати на енергетичній оболонці в явно коваріантній формі, що забезпечує незалежність від імпульсів частинок. Порівняння з відповідним результатом коваріантної теорії збурень Дайсона здійснено за допомогою встановленого зв'язку між старомодною теорією збурень і методом одягаючих перетворень.

Abstract

Applying the unitary clothing transformation method in the model of charged spinless nucleons and neutral mesons interacting via the three-linear Yukawa-type coupling, the expression for the charge shift in the third order in the coupling constant is derived. Being determined off the energy shell, the expression can be reduced on the energy shell to the explicitly covariant form, providing the independence of the particle momenta. Comparison with the corresponding result of the Dyson covariant perturbation theory is performed by establishing the link between the old-fashioned perturbation theory and the clothing approach.

# CONTENTS

# INTRODUCTION

The clothing unitary transformation (UT) method, proposed by Greenberg and Schweber [1] on the basis of deep analysis of the problems of the relativistic quantum field theory (RQFT) performed by Van Hove [2,3], enables to avoid in a natural way several troubles arising in foundations of RQFT (see, e.g., [4] and Refs. therein). One of these troubles is connected with the fact that for the wide class of the field-theoretical models the primary interaction operator, being defined off the energy shell, connects particles which simultaneously stay on their mass shells. Thus, the energies of real particles in the intermediate states are essentially relativistic. Besides, at high enough energies any process has additional inelastic channel associated with the creation and/or destruction of particles, which needs the additional degrees of freedom being explicitly considered. Therefore, the consideration of the relativistic effects off the energy shell and the effects of particle creation/destruction is very topical for the deep understanding of the physics of processes taking place in the meson-nucleon systems in the wide range of energies, including the bound states (see, e.g., [5-7]).

According to [1], the clothing procedure is performed on the total Hamiltonian operator *H* with help of the UT which keeps both the *H*- and *S*-operators (and, thus, all the observables) intact [8] and does the transition from the representation of "bare" particles with primary (unphysical) masses and coupling constants to the representation of "clothed" particles with the observable properties and physical interactions between them. After the elimination from the operators having fundamental physical meaning (such as, the generators of the Poincaré group, *S*-operator, current operator) of several, called "bad", operators which prevent the one-particle states to be the eigenstates of, say, the total Hamiltonian, the latter acquires the so-called "sparse" structure in the Fock space of hadronic states, which enables approximate determination of *H* eigenstates and eigenvalues. It should be noted, that the elimination of "bad" operator does not mean that the latter disappears from the Hamiltonian without any trace. In fact, due to the recursive character of the clothing procedure, the mechanism of some unphysical process (generated by some "bad" operator) appears many times incorporated into the mechanisms of physical processes in higher orders in the coupling constant. In other words, all the infinite variety of "clouds" of virtual quanta, after Van Hove [2,3], appears accumulated into the new creation/destruction operators (for clothed, i.e., physical particles) which are expressed in some quite complex way via the operators of bare particles. Remarkably, that the program of renormalization of physical constants – the masses of particles [9,10] and the coupling constants, as well as the particle wave functions – is automatically performed along with that.

The operators remaining in the Hamiltonian after the clothing appear as the Hermitian operators of physical relativistic interactions between observable particles, which are non-local (relativistic recoil effects are accounted), independent of interaction energy and contain the structures off the energy shell in a natural way.

In this master thesis the procedure of particle clothing is exhibited in several first orders in the coupling constant in the field-theoretical model including the charged spinless nucleon and the neutral meson fields coupled via the Yukawa-type interaction. The expression for the charge shift in the third order in the coupling constant is derived [11-13]. Being determined through the automatic cancellation of the similar operators off the energy shell in the Hamiltonian, this expression consists of two terms. The first one can be reduced to the explicitly covariant form on the energy shell, while the second one turns into zero on the shell. The expression in question on the energy shell is presented through the three-dimensional integrals of Lorentz-scalar combinations of particle four-momenta forming the vertex, which provides the momentum independence of the calculated correction.

To compare our result obtained via the clothing method with the respective one which can be derived using the covariant Dyson-Feynman perturbation theory we do not perform the explicitly covariant calculations. Instead, we establish the general link between the old-fashioned perturbation theory (OFPT) which is, in fact, equivalent to the time-dependent Dyson-Feynman approach and our non-covariant three-dimentional perturbation method. We show that these perturbation approaches appear equivalent on the energy shell. Therefore, we conclude that the expression for the charge shift found in our three-dimensional formalism gives just the other representation for the respective four-dimensional integral of Dyson-Feynman method.

It should be emphasized that the existence of the off energy shell structures in the derived three-dimensional expression for the charge shift plays an important role in the consistent calculations of, e.g., πNN form factors in the problems of nuclear physics.

# CHAPTER 1
# REPRESENTATIONS OF PARTICLES IN QUANTUM FIELD THEORY

1.1. Representation of bare particles with physical masses

Our departure point is the Hamiltonian $H$ expressed in terms of "bare" primary particles:

$$H(\alpha_0) = H_0(\alpha_0) + V(\alpha_0), \tag{1.1}$$

where $H_0(\alpha_0)$ is the free part and $V(\alpha_0)$ is the primary interaction operator. The symbol $\alpha_0$ denotes the set of all creation/destruction operators of bare particles. The latter are characterized by the relativistic energies which, however, depend on the trial (unrenormalized) masses. The interactions between bare particles are determined by the trial coupling constant.

With help of the auxiliary unitary transformation of the operators $\alpha_0$ [4]:

$$\alpha_0 = T\alpha T^\dagger, \tag{1.2}$$

we perform the transition to the new set of creation/destruction operators $\alpha$ which correspond to the particles with physical masses. The latter are now characterized by the relativistic energies in which the physical masses enter.

Taking into account the linear dependence of $H$ (1.1) on $\alpha_0$ and applying the transformation (1.2), we find (details are in [4]):

$$H_0(\alpha_0) = H_F(\alpha) + M_{ren}(\alpha), \tag{1.3}$$
$$V(\alpha_0) = V(\alpha), \tag{1.4}$$

where $M_{ren}$ is the mass counterterm – the operator which determines corrections of particle masses. Note that the appearance of $M_{ren}$ in r.h.s. of (1.3) for the free part of the Hamiltonian leads to the loss of conservation of the number of free particles with new (physical) masses [4].

The substitution of the trial coupling constant by the physical one can be performed via the introduction of the vertex counterterm $V_{ren}$ – the operator which determines the correction of charge. Assuming the linear dependence of $V_0$ on the coupling constant, we put:

$$V_0(g_0) = V(g) + V_{ren}(g_0 - g), \qquad (1.5)$$

where $g_0$ and $g$ are "bare" and observable coupling constants, respectively.

The representation of the Hamiltonian in terms of new creation/destruction operators $\alpha$ is called the representation of bare particles with physical masses [4]:

$$H(\alpha) = H_F(\alpha) + H_I(\alpha) = H_F + V + M_{ren} + V_{ren}, \qquad (1.6)$$

where $H_F$ is the new free part and $H_I$ is the new interaction term.

By definition, the one-particle states generated from "bare" vacuum $\Omega_0$ with help of the operators of bare particles with physical masses $\alpha^\dagger$ are the eigenstates of the free Hamiltonian $H_F$:

$$H_F(\alpha)|\alpha^\dagger \Omega_0\rangle = E|\alpha^\dagger \Omega_0\rangle. \qquad (1.7)$$

At the same time, the existence of the interaction $H_I$ prevents the same one-particle states to be the eigenstates of the total Hamiltonian:

$$H_0(\alpha)\left|\alpha^\dagger\Omega_0\right\rangle \neq E\left|\alpha^\dagger\Omega_0\right\rangle. \tag{1.8}$$

Thus, the solution of the eigenvalue problem for the operator (1.1) in the representation of bare particles with physical masses (1.6) appears impossible.

1.2. Representation of clothed particles

The eigenvalue problem for the Hamiltonian (1.1) in quantum mesodynamics is complicated by the fact that the Fock space of hadronic states has infinite number of dimensions and, therefore, application of the common methods of diagonalization of the Hamiltonian matrix becomes useless. Thus, we could only develop the methods of approximate solution of the stated problem. In particular, with the aim of calculating the amplitudes of the observable processes we can turn to that representation of the Hamiltonian in which it acquires the "sparse" structure enabling approximate finding of its eigenstates and eigenvalues. The transition to the representation of such a kind, according to Greenberg and Schweber [1], can be performed via such reformulation of one and the same initial Hamiltonian under which the latter becomes acting in the new Fock space of the "clothed" hadronic states generated from the clothed (physical) vacuum $\Omega$ with help of the creation/destruction operators of clothed (physical) particles:

$$\alpha_c = W^\dagger \alpha W, \quad WW^\dagger = W^\dagger W = 1, \tag{1.9}$$

where $W$ is the operator of the clothing UT. Transitions of such a kind give unitarily equivalent representations of the Hamiltonian, i.e., they keep the *S*-matrix and all the observables unaltered [8].

By the assumption, in terms of new operators $\alpha_c$, named "clothed", the Hamiltonian satisfies the requirements:

$$H_F(\alpha_c)\left|\alpha_c^\dagger\Omega\right\rangle = E\left|\alpha_c^\dagger\Omega\right\rangle, \tag{1.10}$$

$$H_c(\alpha_c)|\alpha_c^\dagger\Omega\rangle = E|\alpha_c^\dagger\Omega\rangle, \qquad (1.11)$$

so that the new one-particle states are the eigenstates for the free and total Hamiltonians simultaneously, keeping the latter unchanged with that:

$$H(\alpha_0) \equiv H_c(\alpha_c). \qquad (1.12)$$

1.3. Unitary clothing transformation

Let us consider an arbitrary operator $O$ acting in the Fock space of bare hadronic states with physical masses and having polynomial dependence on the creation/destruction operators of the corresponding particles. Applying the unitary transformation (1.9) to this operator and using the well-known formula, we find:

$$\begin{aligned}O(\alpha) = O_c(\alpha_c) &= W(\alpha_c)O(\alpha_c)W^\dagger(\alpha_c) \\ &= e^{R(\alpha_c)}O(\alpha_c)e^{-R(\alpha_c)} = O(\alpha_c) + \sum_{k=1}^{\infty}\frac{1}{k!}\left[R(\alpha_c),O(\alpha_c)\right]^k,\end{aligned} \qquad (1.13)$$

where $W(\alpha_c) = e^{R(\alpha_c)}$ and $R(\alpha_c) = -R^\dagger(\alpha_c)$ is the generator of the UT.

Expression (1.13) uses the following convenient notation for the multiple commutator:

$$[R,O]^k \equiv \underbrace{\left[R,\left[R,...\left[R,O\right]...\right]\right]}_{k}. \qquad (1.14)$$

Inserting the Hamiltonian operator (1.6) instead of the operator $O$ in (1.13), we obtain:

$$H_c(\alpha_c) = H_F(\alpha_c) + H_I(\alpha_c) + \sum_{k=1}^{\infty} \frac{1}{k!}\left[R(\alpha_c), H_F(\alpha_c) + H_I(\alpha_c)\right]^k. \qquad (1.15)$$

The Hamiltonian in the clothed particle representation (1.15) contains operators, called "bad", which prevent the one-clothed-particle states to be the eigenstates of the total Hamiltonian. To keep the property of Hermiticity we assume as "bad" the operators conjugated to them too. Then, the generator $R(\alpha_c)$ should be chosen the way the respective bad operators are absent in the Hamiltonian. The remaining operators are the ones of physical relativistic interactions, i.e., $H_I(\alpha_c)\left|\alpha_c^\dagger \Omega\right\rangle = 0$.

# CHAPTER 2
# FORMALISM OF UNITARY CLOTHING TRANSFORMATIONS

2.1. Types of operators in the Hamiltonian

The operators appearing in the total Hamiltonian during the clothing are conveniently classified the following way[*].

$O_{t,g}^{(n)}$ and $O_{t,b}^{(n)}$ are the "transition" operators of $n$-th order in the coupling constant, which consist of more than three creation/destruction operators for particles of any species. Indices "$g$" and "$b$" differ "good" operators, having the energy shell (possibly, with an energy threshold) and, therefore, corresponding to the observable processes, from "bad" operators, having no such shell and, therefore, corresponding to virtual processes. Note that all bad operators are virtual but not all virtual operators are bad. Any transition operator has a unique decomposition $O_t^{(n)} = O_{t,g}^{(n)} + O_{t,b}^{(n)}$.

$O_{M_r}^{(n)}$ and $O_{V_r}^{(n)}$ are the operators of "mass" and "vertex" types of $n$-th order in the coupling constant, which replicate the operator structures of mass and vertex counterterms $M_{ren}$ and $V_{ren}$, respectively. Note that the operator $O_{M_r}^{(n)}$ can be both good "$g$" and bad "$b$". Along with that, in case the primary interaction operator $V$ totally consists of bad operators, the operator $O_{V_r}^{(n)}$ is always bad. It is assumed that these operators have the following series in the coupling constant:

$$M_{ren} = \sum_{k=1}^{\infty} M_{ren}^{(2k)}, \qquad (2.1)$$

$$V_{ren} = \sum_{k=1}^{\infty} V_{ren}^{(2k+1)}, \qquad (2.2)$$

---

[*] Being more specific but not restricting the exposition, we assume the operator $V$ in (1.1) to be a three-linear one of the Yukawa-type.

thus, the analogous expansions are expected for the corrections of physical constants.

The algebra of operators arising in the Hamiltonian during the clothing is presented in Table 1.

For the model of interacting nucleons and mesons, in which $b^\dagger(b)$ and $d^\dagger(d)$ are the creation (destruction) operators for nucleons and antinucleons and $a^\dagger(a)$ are the creation (destruction) operators for mesons, the operator $b^\dagger b a^\dagger a^\dagger$ is of bad transition type, while $b^\dagger b^\dagger bb$ is of good transition one corresponding to the observable process of nucleon-nucleon scattering $NN \to NN$, $b^\dagger b$ and $b^\dagger d^\dagger$ are good and bad operators of mass type, respectively, finally $b^\dagger b a^\dagger$ and $b^\dagger d^\dagger a^\dagger$ are the vertex type operators.

Table 1.

Algebra of operators of different types in the Hamiltonian. Symbols $t,g$ and $t,b$ mark good and bad transition operators, respectively, $M_{r,g}$ and $M_{r,b}$ denote good and bad mass-type operators, $V_r$ are the vertex-like operators.

| Types of operators | Commutators of operators of the type | | | | |
|---|---|---|---|---|---|
| | $t,g$ | $t,b$ | $M_{r,g}$ | $M_{r,b}$ | $V_r$ |
| $t,b$ | $t + M_{r,b} + V_r$ | all types | $t,b$ | all types | all types |
| $V_r$ | all types | all types | $V_r$ | $V_r$ | $t + M_r$ |
| $M_{r,b}$ | $t + M_{r,b} + V_r$ | all types | $M_{r,b}$ | $M_{r,g}$ | $V_r$ |

2.2. First clothing transformation

After the clothing, the total Hamiltonian in the form (1.15) contains bad terms of the first order in the coupling constant (the operator $V$) and the ones of the second and higher orders if the expansions (2.1) and (2.2) are taken into account. Thus, the generator $R$ which repeats the structure of bad operators should be chosen as the similar expansion in the coupling constant:

$$R = \sum_{k=1}^{\infty} R^{(k)}. \tag{2.3}$$

Using the formulae (2.1)-(2.3), the Hamiltonian (1.15) can be given as:

$$H_c(\alpha_c) = H_F + \left[R^{(1)}, H_F\right] + V + M_{ren}^{(2)} + \left[R^{(1)}, V\right] + \left[R^{(2)}, H_F\right]$$

$$+ V_{ren}^{(3)} + \frac{1}{2}\left[R^{(1)}, V\right]^2 + \left[R^{(1)}, M_{ren}^{(2)}\right] + \left[R^{(3)}, H_F\right] + ..., \tag{2.4}$$

where the terms up to the third order in $g$ are explicitly shown.

For the wide class of field-theoretical models the primary interaction operator $V$ consists totally of the bad operators of the first order $H_b^{(1)} \equiv V$. The condition for them to be absent in (2.4)

$$H_b^{(1)} + \left[R^{(1)}, H_F\right] = 0, \tag{2.5}$$

is the equation determining $R^{(1)}$. If the equation (2.5) is solvable then the Hamiltonian (2.4) acquires the form free from bad terms of the first order:

$$H_c(\alpha_c) = H_F(\alpha_c) + \left[R^{(2)}, H_F\right] + \frac{1}{2}\left[R^{(1)}, V\right] + M_{ren}^{(2)}$$

$$+ \left[R^{(3)}, H_F\right] + \frac{1}{3}\left[R^{(1)}, V\right]^2 + \left[R^{(1)}, M_{ren}^{(2)}\right] + V_{ren}^{(3)} + .... \tag{2.6}$$

So far, by the assumption, the generator $R^{(1)}$ repeats the structure of bad terms of the first order, i.e. $V$, then, using the Table 1, we find that the r.h.s. of (2.6) contains bad terms of the second order which stem from $\left[R^{(1)},V\right]$ and $M_{ren}^{(2)}$. Before eliminating these operators with the second clothing transformation we need to separate them first in the explicit form. To do that we classify the operators of the mentioned order in the Hamiltonian and collect similar terms. According to the Table 1, the commutator $\left[R^{(1)},V\right]$ consists of good transition operators corresponding to the physical processes in the second order, bad transition operators and also good and bad mass-type operators:

$$\left[R^{(1)},V\right]=\left[R^{(1)},V\right]_{t,g}+\left[R^{(1)},V\right]_{t,b}+\left[R^{(1)},V\right]_{M_r,g}+\left[R^{(1)},V\right]_{M_r,b}. \tag{2.7}$$

By definition, the mass counterterm consists of good and bad parts: $M_{ren}^{(2)}=M_{ren,g}^{(2)}+M_{ren,b}^{(2)}$. Therefore, putting

$$M_{ren,g}^{(2)}+\frac{1}{2}\left[R^{(1)},V\right]_{M_r,g}=0, \tag{2.8}$$

we eliminate from the Hamiltonian good mass-type operators of the second order and automatically fix the values for the mass shifts [9,10]. At the same time, collecting bad mass-type operators after the values for mass shifts are fixed, we find that in general the condition analogous to (2.8) is not satisfied [10]:

$$M_{ren,b}^{(2)}+\frac{1}{2}\left[R^{(1)},V\right]_{M_r,b}\equiv M_{ren,b,rest}\neq 0. \tag{2.9}$$

Thus, in the Hamiltonian

$$H_c(\alpha_c) = H_F(\alpha_c) + H_g^{(2)} + \left[R^{(2)}, H_F\right] + H_b^{(2)}$$
$$+ \left[R^{(3)}, H_F\right] + \frac{1}{3}\left[R^{(1)}, V\right]^2 + \left[R^{(1)}, M_{ren}^{(2)}\right] + V_{ren}^{(3)} + \ldots, \quad (2.10)$$

we are able to separate good operators

$$H_g^{(2)} = \frac{1}{2}\left[R^{(1)}, V\right]_{t,g}, \quad (2.11)$$

corresponding to the observable processes in the second order and bad ones

$$H_b^{(2)} = \frac{1}{2}\left[R^{(1)}, V\right]_{t,b} + M_{ren,b,rest}^{(2)}, \quad (2.12)$$

which must be removed via the second clothing UT.

2.3. Second clothing transformation

The generator $R^{(2)}$ should be determined from the condition of absence in the r.h.s. of (2.10) of bad operators of the second order:

$$H_b^{(2)} + \left[R^{(2)}, H_F\right] = 0. \quad (2.13)$$

If this equation is solvable as well as the analogous equation (2.5) then after the second clothing the total Hamiltonian contains only good operators corresponding to the observable processes in the second order:

$$H_c(\alpha_c) = H_F(\alpha_c) + H_g^{(2)}$$
$$+ \left[R^{(3)}, H_F\right] + \frac{1}{3}\left[R^{(1)}, V\right]^2 + \left[R^{(1)}, M_{ren}^{(2)}\right] + V_{ren}^{(3)}$$

$$+\left[R^{(4)}, H_F\right] + M_{ren}^{(4)} + T^{(4)} + \ldots, \tag{2.14}$$

where the fourth order operators are additionally explicitly shown:

$$T^{(4)} = \frac{1}{8}\left[R^{(1)}, V\right]^3 + \left[R^{(2)}, H_g^{(2)}\right]$$

$$+ \frac{1}{2}\left[R^{(2)}, H_b^{(2)}\right] + \frac{1}{2}\left[R^{(1)}, M_{ren}^{(2)}\right]^2 + \left[R^{(1)}, V_{ren}^{(3)}\right]. \tag{2.15}$$

The third order operator $\frac{1}{3}\left[R^{(1)}, V\right]^2 + \left[R^{(1)}, M_{ren}^{(2)}\right] + V_{ren}^{(3)}$ in (2.14) contains good and bad terms. According to the Table 1, the commutator $\left[R^{(1)}, V\right]^2$ can be split as follows: $\left[R^{(1)}, V\right]^2 = \left[R^{(1)}, V\right]^2_{t,g} + \left[R^{(1)}, V\right]^2_{t,b} + \left[R^{(1)}, V\right]^2_{V_r}$. In its turn, the commutator $\left[R^{(1)}, M_{ren}^{(2)}\right] = \left[R^{(1)}, M_{ren}^{(2)}\right]_{V_r}$ consists of the vertex-like operators. As shown in [11-16], in a general case, collecting the similar terms of the vertex type gives:

$$\frac{1}{3}\left[R^{(1)}, V\right]^2_{V_r} + \left[R^{(1)}, M_{ren}^{(2)}\right]_{V_r} + V_{ren}^{(3)} \equiv V_{ren,rest}^{(3)} \neq 0, \tag{2.16}$$

which, however, does not prevent fixing the charge correction in the third order.

Therefore, after the second clothing the Hamiltonian acquires the form:

$$H_c(\alpha_c) = H_F(\alpha_c) + H_g^{(2)} + H_g^{(3)} + \left[R^{(3)}, H_F\right] + H_b^{(3)}$$

$$+ \left[R^{(4)}, H_F\right] + M_{ren}^{(4)} + T^{(4)} + \ldots, \tag{2.17}$$

where we explicitly extract good operators

$$H_g^{(3)} = \frac{1}{3}\left[R^{(1)}, V\right]_{t,g}^2, \tag{2.18}$$

corresponding to the observable processes in the third order and the bad ones:

$$H_b^{(3)} = \frac{1}{3}\left[R^{(1)}, V\right]_{t,b}^2 + V_{ren,rest}^{(3)}, \tag{2.19}$$

which must be removed via the third order clothing UT.

2.4. Subsequent clothing transformations

In order to clean the total Hamiltonian (2.17) out of bad terms of the third order it is now sufficient to require:

$$H_b^{(3)} + \left[R^{(3)}, H_F\right] = 0, \tag{2.20}$$

and repeat all analytical actions described in subsections 2.2 and 2.3.

Further, after elimination of the bad operators of the third order with use of the correlation (2.20), the formal procedure of extracting and collecting the mass-type operators in the fourth order and fixing the mass shifts in that order in the coupling constant looks the same as the procedure of determining mass shifts in the second order:

$$M_{ren,g}^{(4)} + T_{M_r,g}^{(4)} = 0, \tag{2.21}$$

$$M_{ren,b}^{(4)} + T_{M_r,b}^{(4)} \equiv M_{ren,b,rest}^{(4)} \neq 0, \tag{2.22}$$

where the operators $T^{(4)}_{M_r,g}$ and $T^{(4)}_{M_r,b}$ include good and bad mass-type operators comprising the operator $T^{(4)}$ (2.15). Now one can extract bad terms of the fourth order in the Hamiltonian:

$$H^{(4)}_b = M^{(4)}_{ren,b,rest} + T^{(4)}_{t,b}, \qquad (2.23)$$

where $T^{(4)}_{t,b}$ are the bad transition operators in $T^{(4)}$ and remove them via the fourth clothing UT by selecting the generator $R^{(4)}$ in accordance with the condition:

$$H^{(4)}_b + \left[ R^{(4)}, H_F \right] = 0. \qquad (2.24)$$

Proceeding with extracting and collecting bad operators of higher orders, we fix the charge shift in the fifth order as a byproduct, similarly to the determination of that value in the third order:

$$V^{(5)}_{ren} + T^{(5)}_{V_r} \equiv V^{(5)}_{ren,rest} \neq 0, \qquad (2.25)$$

where:

$$T^{(5)} = \frac{1}{2}\left[ R^{(3)}, \left[ R^{(1)}, V \right] \right]_{t,g} + \frac{1}{3}\left[ R^{(2)}, \left[ R^{(1)}, V \right]^2 \right]_{t,g} + \frac{1}{30}\left[ R^{(1)}, V \right]^4$$
$$+ \left[ R^{(1)}, M^{(4)}_{ren} \right] + \frac{1}{3!}\left[ R^{(1)}, M^{(2)}_{ren} \right]^3 + \left[ R^{(2)}, V^{(3)}_{ren,rest} \right] + \frac{1}{2}\left[ R^{(1)}, V^{(3)}_{ren} \right]^2. \qquad (2.26)$$

And so forth.

The exposed procedure of manipulating with one and the same initial Hamiltonian (1.1) with the aim of obeying the conditions (1.10) and (1.11) via the extraction of bad terms, collecting the similar ones and eliminating them from the Hamiltonian can be generalized on the case of arbitrary order in the coupling constant. Really, comparing the formulae (2.5), (2.13), (2.20) and (2.24), we note that the generator of the *n*-order clothing UT $R^{(n)}$ should be determined from the condition:

$$H_b^{(n)} + \left[ R^{(n)}, H_F \right] = 0, \quad n = 1, 2, ... \tag{2.27}$$

where $H_b^{(n)}$ is the *n*-order bad term. However, one should bear in mind that in general $R^{(n)}$ can be determined from the solution of equation (2.27) only after all the generators $R^{(k)}$, $k = 1, 2, ..., n-1$, are fixed because only in that case the explicit form of the operator $H_b^{(n)}$ is known. In other words, the exposed procedure, by the construction, has a recursive character. That is why the structure of the Hamiltonian in the *n*-order in the coupling constant cannot be explicitly known before the corrections to the physical constants are fixed up to the *n*-order and the bad terms up to that order are extracted and removed.

2.5. The clothing method vs. other approaches using unitary transformations

There exists a variety of different approaches in the RQFT which use the formalism of UT's (see, e.g., [9] and Refs. therein). Let us perform a comparative analysis of physical and mathematical foundations of those of them the application of which enables to construct the Hermitian relativistic operators of few-body interactions, being energy independent, having off energy shell structures and originating on one and the same physical background.

Okubo [17] proposes to split the initial full Fock space of hadronic states into the subspaces (sectors), depending on the described physical process. For instance, in order to calculate the nucleon mass shift the one-nucleon subspace is separated out of full space [7,18], to construct the operator for the NN→NN scattering the two-nucleon subspace is additionally extracted from remaining part of full space [5,6,18-21], to analyze the process πN→πN the remaining space is further divided and the pion-nucleon sector is separated, and so forth. As a result, the Hamiltonian operator appears reduced to the block-diagonal form. Unfortunately, the described approach in that formulation leaves no possibility of considering all the physical processes observed in nature. In particular, one of the most important and interesting from the fundamental viewpoint processes of the new particle production at threshold (e.g., NN↔πNN) needs for their treatment additional intellectual efforts. On should note, that the considered method of splitting the space of states appears as the substitution of the initial Hamiltonian by the other one which is unitarily equivalent to the former [22]. The generators of the corresponding subsequent transformations are, in fact, chosen due to the condition that the new Hamiltonian is free from the operators associated with physical processes with particle redistribution (transitions between sectors with different numbers of particles).

Heitler [23] has obtained the operator differential equation to which the operator of his UT must obey in the interaction picture, if one focuses on the elimination from the Hamiltonian of the operators corresponding to the "virtual" processes (i.e., the processes which have no energy shell) that bring no contribution to the $S$-matrix. Lee and Sato with collaborators [24,25] have shown that the elimination of "virtual" operators form the Hamiltonian can be performed in the Schrödinger picture too. As in case of Okubo transformation, the Heitler-Lee-Sato transformation appears the substitution of the initial Hamiltonian by the new one which is unitarily equivalent to the former.

In the Heitler-Lee-Sato approach one succeeds to avoid the most substantial drawbacks peculiar to the Okubo method. In particular, the operators of physical processes with particle redistribution are calculated alongside the operators of other physical processes. It is important to note, that the operators of real processes act in the full Fock space of hadronic states and originate from one and the same initial Hamiltonian without any additional assumptions of physical or mathematical character. At the same time one notes the "hidden" dependence of the approach and the derived operators of physical processes on the interaction energy. Really, for instance, the physical process $\pi\pi \to N\bar{N}$ has a natural energy threshold below which it is virtual. Consequently, depending on the energy of interaction, the operator of the given process must or must not be removed from the Hamiltonian by some UT.

Analyzing the physical and mathematical foundations of the methods by Okubo, Heitler-Lee-Sato and Greenberg-Schweber, one comes to the conclusion that in the first two approaches the particles stay bare in spite of the UT's performed on them, and only the latter one operates with really clothed (i.e., including the virtual effects, after Van Hove) particles. Along with that, the explicit analytical expressions for the operators of physical interactions found with help of the described methods in the first non-vanishing orders in the coupling constant appear algebraically the same while having quit different physical content.

From the mathematical viewpoint, the strongest condition imposed upon the operators eliminated from the Hamiltonian is, obviously, the Okubo's one. So far the "virtual" processes can be generated not only by bad, after Greenberg and Schweber, but also by good operators in the Hamiltonian the Heitler-Lee-Sato's requirement is weaker than the Okubo's one but stronger than the adopted by us Greenberg-Schweber's condition. In other words, the class of operators defined as "bad" is narrower that the class of "virtual" operators while the latter appears narrower than the class of transition operators between sectors in Fock space. One also emphasizes that in the Greenberg-Schweber approach the program of renormalization of physical constants is performed automatically as a byproduct of the clothing procedure along with the construction of the operators of physical interactions.

From the physical viewpoint, the mentioned distinctions will obviously manifest themselves in the details of mechanisms via which the physical processes between physical particles take place. The appearance of these distinctions in the results for the observed values should be expected yet in the fourth order in the coupling constant because namely in that order their appear first corrections connected with the application of the forthcoming UT's.

# CHAPTER 3
# VERTEX RENORMALIZATION
# IN CLOTHED PARTICLE REPRESENTATION

## 3.1. The field-theoretical model

Let us implement the developed technique in the field-theoretical model of scalar meson and charged spinless nucleon fields interacting via the three-linear Yukawa-type coupling. The total Hamiltonian operator in this model has the following form:

$$H(\alpha) = H_F + V + M_{ren} + V_{ren}, \tag{3.1}$$

$$H_F = \int d^3q E_q \left(b_{\vec{q}}^\dagger b_{\vec{q}} + d_{\vec{q}}^\dagger d_{\vec{q}}\right) + \int d^3k \omega_k a_{\vec{k}}^\dagger a_{\vec{k}}, \tag{3.2}$$

$$V = \int d^3k \, \hat{V}^{\vec{k}} \, a_{\vec{k}}^\dagger + H.c., \tag{3.3}$$

$$\hat{V}^{\vec{k}} = \int d^3p \, d^3q : F_i^{\vec{p}\dagger} V_{i,j}^{\vec{k}}(\vec{p},\vec{q}) F_j^{\vec{q}} :, \qquad \hat{V}^{\vec{k}\dagger} = \hat{V}^{-\vec{k}}, \tag{3.4}$$

$$V_{i,j}^{\vec{k}}(\vec{p},\vec{q}) = -\frac{g}{(2\pi)^{3/2}} \frac{\delta(\vec{p}-\vec{q}+\vec{k})}{(8E_p E_q \omega_k)^{1/2}} \varepsilon_{i,j},$$

$$\varepsilon_{i,j} = \begin{pmatrix} 1 & 1 \\ 1 & 1 \end{pmatrix}, \quad i,j = 1, 2. \tag{3.5}$$

Here we adopt the following denotation for the nucleon creation/destruction operators:

$$F_i^{\vec{q}} = \begin{pmatrix} b_{\vec{q}} \\ d_{-\vec{q}}^\dagger \end{pmatrix}, \; F_i^{\vec{q}\dagger} = \begin{pmatrix} b_{\vec{q}}^\dagger & d_{-\vec{q}} \end{pmatrix}, \tag{3.6}$$

where $b_{\vec{q}}^{\dagger}\left(b_{\vec{q}}\right)$ and $d_{\vec{q}}^{\dagger}\left(d_{\vec{q}}\right)$ are the creation (destruction) operators of nucleons and antinucleons with the momentum $\vec{q}$, respectively. Operators $F_i^{\vec{q}\dagger}$ and $F_i^{\vec{q}}$ satisfy the following commutation relations

$$\left[F_i^{\vec{p}}, F_j^{\vec{q}\dagger}\right] = (-1)^{i+1}\delta_{ij}\delta(\vec{p}-\vec{q}), \qquad i,j = 1, 2, \tag{3.7}$$

which are generated by the usual commutation relations of Bose operators: $\left[b_{\vec{p}}, b_{\vec{q}}^{\dagger}\right] = \delta(\vec{p}-\vec{q})$ and $\left[d_{\vec{p}}, d_{\vec{q}}^{\dagger}\right] = \delta(\vec{p}-\vec{q})$.

Creation (destruction) operators of mesons with the momentum $\vec{k}$, $a_{\vec{k}}^{\dagger}\left(a_{\vec{k}}\right)$ satisfy the relation:

$$\left[a_{\vec{k}}, a_{\vec{k}'}^{\dagger}\right] = \delta(\vec{k}-\vec{k}'). \tag{3.8}$$

The mass counterterm consists of the mesonic and nucleonic parts which determine the meson $\delta\mu^2 = \mu_0^2 - \mu^2$ and nucleon $\delta m^2 = m_0^2 - m^2$ mass shifts, respectively:

$$M_{ren} = M_{ren,mes} + M_{ren,nucl}, \tag{3.9}$$

$$M_{ren,mes} = \delta\mu^2 \int \frac{d^3k_1 d^3k_2}{4\sqrt{\omega_{k_1}\omega_{k_2}}} \delta(\vec{k}_1 - \vec{k}_2)\left(a_{\vec{k}_1}^{\dagger}a_{\vec{k}_2} + a_{\vec{k}_1}^{\dagger}a_{-\vec{k}_2}^{\dagger}\right) + H.c., \tag{3.10}$$

$$M_{ren,nucl} = \delta m^2 \int d^3q\, d^3p : F_i^{\vec{p}\dagger} M_{i,j}(\vec{p},\vec{q}) F_j^{\vec{q}} : + H.c., \tag{3.11}$$

$$M^{i,j}(\vec{p},\vec{q}) = \frac{\delta(\vec{p}-\vec{q})}{8\sqrt{E_q E_p}}\varepsilon_{i,j}, \quad i,j = 1, 2. \tag{3.12}$$

Here $E_p = \sqrt{m^2 + \vec{p}^2}$ is the energy of a physical nucleon with the momentum $\vec{p}$, $\omega_k = \sqrt{\mu^2 + \vec{k}^2}$ is the energy of a physical meson with the momentum $\vec{k}$, $m$ and $\mu$ are the corresponding physical masses of particles, while $m_0$ and $\mu_0$ are their bare counterparts.

The vertex counterterm $V_{ren} \equiv V(g_0) - V(g)$ determines the correction of the charge value $\delta g = g_0 - g$ where $g_0$ is the bare coupling constant and $g$ is its observable counterpart:

$$V_{ren} = -\frac{\delta g}{(2\pi)^{3/2}} \int \frac{d^3k\, d^3p\, d^3q}{(8E_p E_q \omega_k)^{1/2}} \delta(\vec{p}-\vec{q}+\vec{k}) : F_{\vec{p}}^{i\dagger} \varepsilon_{i,j} F_{\vec{q}}^{j} : a_{\vec{k}}^{\dagger} + H.c.. \qquad (3.13)$$

3.2. Mass renormalization

In the model under consideration, bad terms $H_b^{(1)}$ of the first order coincide with the interaction operators $V$ (3.3). Assuming the generator of first clothing $R^{(1)}$ replicating the structure of $H_b^{(1)}$, the solution of Eq. (2.5) acquires the form:

$$R^{(1)} = \int d^3k\, \hat{R}^{\vec{k}} a_{\vec{k}}^{\dagger} - H.c., \qquad (3.14)$$

$$\hat{R}^{\vec{k}} = \int d^3p\, d^3q : F_i^{\vec{p}\dagger} R_{i,j}^{\vec{k}}(\vec{p},\vec{q}) F_j^{\vec{q}} : \qquad (3.15)$$

$$R_{i,j}^{\vec{k}}(\vec{p},\vec{q}) = -V_{i,j}^{\vec{k}}(\vec{p},\vec{q}) \frac{1}{(-1)^{i+1} E_p - (-1)^{j+1} E_q + \omega_k}, \quad i,j = 1, 2. \qquad (3.16)$$

Such choice of the generator $R^{(1)}$ allows cleaning of the Hamiltonian (3.1) out of $g^1$-order bad terms.

Using the recipe of the paragraph 2.3, to implement the second clothing UT we have to collect operators in the $g^2$ order. The expression for the commutator $\left[R^{(1)}, V\right]$ of the second order is as follows:

$$\left[R^{(1)}, V\right] = \int d^3k_1\, d^3k_2$$

$$\times \left[ a^\dagger_{\vec{k}_1} a^\dagger_{\vec{k}_2} \left[ \hat{R}^{\vec{k}_2}, \hat{V}^{\vec{k}_1} \right] + a^\dagger_{\vec{k}_2} a_{\vec{k}_1} \left[ \hat{R}^{\vec{k}_2}, \hat{V}^{-\vec{k}_1} \right] - \delta(\vec{k}_1 - \vec{k}_2) \hat{R}^{\vec{k}_2} \hat{V}^{-\vec{k}_1} \right] + H.c.. \quad (3.17)$$

Using the Table 1 and the normal ordering, one can represent the commutator $\left[ R^{(1)}, V \right]$ as the sum of operators of different types, namely, good and bad transition operators and meson and nucleon mass-like operators, respectively:

$$\left[ R^{(1)}, V \right] = \left[ R^{(1)}, V \right]_{t,g} + \left[ R^{(1)}, V \right]_{t,b} + \left[ R^{(1)}, V \right]_{M_r, mes} + \left[ R^{(1)}, V \right]_{M_r, nucl}. \quad (3.18)$$

Actually, besides these operators the commutator $\left[ R^{(1)}, V \right]$ contains the constant value providing the vacuum energy renormalization in the second order in the coupling constant. However, this problem is out of scope of the present work.

In the partition (3.18), the operators comprising $\left[ R^{(1)}, V \right]_{t,g}$ are the ones associated with the observable processes in the second order in coupling constant (see also [9]):

$$\frac{1}{2} \left[ R^{(1)}, V \right]_{t,g} = H^{(2)}(\text{NN} \to \text{NN}) + H^{(2)}(\text{N}\bar{\text{N}} \to \text{N}\bar{\text{N}}) + H^{(2)}(\bar{\text{N}}\bar{\text{N}} \to \bar{\text{N}}\bar{\text{N}})$$

$$+ H^{(2)}(\pi \text{N} \to \pi \text{N}) + H^{(2)}(\pi \bar{\text{N}} \to \pi \bar{\text{N}}) + H^{(2)}(\text{N}\bar{\text{N}} \leftrightarrow \pi\pi), \quad (3.19)$$

where we adopt the transparent denotations for nucleon (N), antinucleon ($\bar{\text{N}}$) and meson ($\pi$).

Operators of nucleon (antinucleon) – nucleon (antinucleon) interactions in the second order in $g$ look as follows:

$$H^{(2)}(\text{NN} \to \text{NN}) = \int d^3 p_2 d^3 q_2 d^3 q_1 d^3 p_1$$

$$\times V^{(\mathrm{NN}\to\mathrm{NN})}\left(\vec{p}_2,\vec{q}_1;\vec{q}_2,\vec{p}_1\right)F_1^{\vec{p}_2\,\dagger}F_1^{\vec{q}_1\,\dagger}F_1^{\vec{q}_2}F_1^{\vec{p}_1}+H.c.,$$

$$V^{(\mathrm{NN}\to\mathrm{NN})}\left(\vec{p}_2,\vec{q}_1;\vec{q}_2,\vec{p}_1\right)=-\frac{1}{2}\int d^3k_1 R_{1,1}^{\vec{k}_1}(\vec{p}_2,\vec{q}_2)V_{1,1}^{\vec{k}_1}(\vec{p}_1,\vec{q}_1); \qquad (3.20)$$

$$H^{(2)}\left(\overline{\mathrm{N}}\overline{\mathrm{N}}\to\overline{\mathrm{N}}\overline{\mathrm{N}}\right)=\int d^3p_2 d^3q_2 d^3q_1 d^3p_1$$

$$\times V^{(\overline{\mathrm{N}}\overline{\mathrm{N}}\to\overline{\mathrm{N}}\overline{\mathrm{N}})}\left(\vec{q}_2,\vec{p}_1;\vec{q}_1,\vec{p}_2\right)F_2^{\vec{q}_2}F_2^{\vec{p}_1}F_2^{\vec{q}_1\,\dagger}F_2^{\vec{p}_2\,\dagger}+H.c.,$$

$$V^{(\overline{\mathrm{N}}\overline{\mathrm{N}}\to\overline{\mathrm{N}}\overline{\mathrm{N}})}\left(\vec{q}_2,\vec{p}_1;\vec{q}_1,\vec{p}_2\right)=-\frac{1}{2}\int d^3k_1 R_{2,2}^{\vec{k}_1}(\vec{p}_2,\vec{q}_2)V_{2,2}^{\vec{k}_1}(\vec{p}_1,\vec{q}_1); \qquad (3.21)$$

$$H^{(2)}\left(\mathrm{N}\overline{\mathrm{N}}\to\mathrm{N}\overline{\mathrm{N}}\right)=\int d^3p_2 d^3q_2 d^3q_1 d^3p_1$$

$$\times V^{(\mathrm{N}\overline{\mathrm{N}}\to\mathrm{N}\overline{\mathrm{N}})}\left(\vec{p}_2,\vec{p}_1;\vec{q}_2,\vec{q}_1\right)F_1^{\vec{p}_2\,\dagger}F_2^{\vec{p}_1}F_1^{\vec{q}_2}F_2^{\vec{q}_1\,\dagger},$$

$$V^{(\mathrm{N}\overline{\mathrm{N}}\to\mathrm{N}\overline{\mathrm{N}})}\left(\vec{p}_2,\vec{p}_1;\vec{q}_2,\vec{q}_1\right)=$$

$$-\frac{1}{2}\int d^3k_1\left[R_{1,1}^{\vec{k}_1}(\vec{p}_2,\vec{q}_2)V_{2,2}^{\vec{k}_1}(\vec{p}_1,\vec{q}_1)+R_{2,2}^{\vec{k}_1}(\vec{q}_1,\vec{p}_1)V_{1,1}^{\vec{k}_1}(\vec{q}_2,\vec{p}_2)\right]$$

$$-\frac{1}{2}\int d^3k_1\left[R_{1,2}^{\vec{k}_1}(\vec{p}_2,\vec{p}_1)V_{1,2}^{\vec{k}_1}(\vec{q}_2,\vec{q}_1)+R_{2,1}^{\vec{k}_1}(\vec{q}_1,\vec{q}_2)V_{2,1}^{\vec{k}_1}(\vec{p}_2,\vec{p}_1)\right]. \qquad (3.22)$$

Operators of the processes of creation (destruction) of two mesons from (into) the nucleon-antinucleon pair are:

$$H^{(2)}\left(\mathrm{N}\overline{\mathrm{N}}\leftrightarrow\pi\pi\right)=\int d^3k_1 d^3k_2 d^3q d^3p$$

$$\times V^{(\mathrm{N}\overline{\mathrm{N}}\leftrightarrow\pi\pi)}\left(\vec{k}_1,\vec{k}_2;\vec{p},\vec{q}\right)F_2^{\vec{p}}F_2^{\vec{q}\,\dagger}a_{\vec{k}_1}^{\dagger}a_{\vec{k}_2}^{\dagger}+H.c.,$$

$$V^{(\mathrm{N}\overline{\mathrm{N}}\leftrightarrow\pi\pi)}\left(\vec{k}_1,\vec{k}_2;\vec{p},\vec{q}\right)=$$

$$=\frac{1}{2}(-1)^{i+1}\int d^3q_1\left[R_{2,i}^{\vec{k}_1}(\vec{q},\vec{q}_1)V_{i,2}^{\vec{k}_2}(\vec{q}_1,\vec{p})-V_{2,i}^{\vec{k}_1}(\vec{q},\vec{q}_1)R_{i,2}^{\vec{k}_2}(\vec{q}_1,\vec{p})\right]. \qquad (3.23)$$

Operators of the meson – nucleon (antinucleon) scattering are given by the expressions:

$$H^{(2)}(\pi N \to \pi N) = \int d^3p d^3k_1 d^3p d^3k_2$$

$$\times V^{(\pi N \to \pi N)}(\vec{p},\vec{k}_1;\vec{q},\vec{k}_2) F_1^{\vec{p}\dagger} F_1^{\vec{q}} a^\dagger_{\vec{k}_1} a_{\vec{k}_2} + H.c.,$$

$$V^{(\pi N \to \pi N)}(\vec{p},\vec{k}_1;\vec{q},\vec{k}_2) =$$

$$= \frac{1}{2}(-1)^{i+1} \int d^3q_1 \left[ R_{1,i}^{\vec{k}_2}(\vec{p},\vec{q}_1) V_{i,1}^{-\vec{k}_1}(\vec{q}_1,\vec{q}) - R_{1,i}^{\vec{k}_1}(\vec{p},\vec{q}_1) V_{i,1}^{\vec{k}_2}(\vec{q}_1,\vec{q}) \right]; \quad (3.24)$$

$$H^{(2)}(\pi \bar{N} \to \pi \bar{N}) = \int d^3p_1 d^3k_1 d^3p_2 d^3k_2$$

$$\times V^{(\pi \bar{N} \to \pi \bar{N})}(\vec{p},\vec{k}_1;\vec{q},\vec{k}_2) F_2^{\vec{p}} F_2^{\vec{q}\dagger} a^\dagger_{\vec{k}_1} a_{\vec{k}_2} + H.c.,$$

$$V^{(\pi \bar{N} \to \pi \bar{N})}(\vec{p},\vec{k}_1;\vec{q},\vec{k}_2) =$$

$$= \frac{1}{2}(-1)^{i+1} \int d^3q_1 \left[ R_{2,i}^{\vec{k}_2}(\vec{p},\vec{q}_1) V_{i,2}^{-\vec{k}_1}(\vec{q}_1,\vec{q}) - R_{2,i}^{\vec{k}_1}(\vec{p},\vec{q}_1) V_{i,2}^{\vec{k}_2}(\vec{q}_1,\vec{q}) \right], \quad (3.25)$$

where the summation over the repeating indices is assumed.

The commutator $\left[ R^{(1)}, V \right]_{t,b}$ consists of bad transition operators (e.g., $b^\dagger b^\dagger d^\dagger b$, $b^\dagger b a^\dagger a^\dagger$) which are to be eliminated from the Hamiltonian via the second clothing UT.

Further, collecting the mass-like operators of the second order in $g$, one can fix the mass shifts. Let us write down the mass-like operators of the second order in the coupling constant contained in the commutator $\left[ R^{(1)}, V \right]$:

$$\left[ R^{(1)}, V \right]_{M_r,mes} = \int d^3k_1 d^3k_2 d^3q d^3p \left( a^\dagger_{-\vec{k}_1} a^\dagger_{\vec{k}_2} + a^\dagger_{\vec{k}_2} a_{\vec{k}_1} \right)$$

$$\times \left[ R_{2,1}^{\vec{k}_2}(\vec{p},\vec{q}) V_{1,2}^{-\vec{k}_1}(\vec{q},\vec{p}) - V_{2,1}^{-\vec{k}_1}(\vec{q},\vec{p}) R_{1,2}^{\vec{k}_2}(\vec{p},\vec{q}) \right] + H.c. \quad (3.26)$$

$$\left[R^{(1)}, V\right]_{M_r, nucl} = \int d^3 p_1 d^3 p_2\, O_{i_2, i_1}(\vec{p}_2, \vec{p}_1) F_{i_2}^{\vec{p}_2\,\dagger} F_{i_1}^{\vec{p}_1} + H.c., \quad (3.27)$$

$$O_{i_2, i_1}(\vec{p}_2, \vec{p}_1) = -\int d^3 q\, d^3 k \left[ R_{i_2,1}^{\vec{k}}(\vec{p}_2, \vec{q}) V_{1,i_1}^{-\vec{k}}(\vec{q}, \vec{p}_1) + R_{2,i_1}^{-\vec{k}}(\vec{q}, \vec{p}_1) V_{i_2,2}^{\vec{k}}(\vec{p}_2, \vec{q}) \right]. (3.28)$$

Note that the kernel (3.28) of the operator (3.27) is independent of the indices of nucleon creation/destruction operators and can be expressed through the matrix $\varepsilon_{i,j}$:

$$\int d^3 p_1 d^3 p_2\, O_{i_2, i_1}(\vec{p}_2, \vec{p}_1) F_{i_2}^{\vec{p}_2\,\dagger} F_{i_1}^{\vec{p}_1} + H.c.$$

$$\equiv \int d^3 p_1 d^3 p_2 \left[ O_{i_2, i_1}(\vec{p}_2, \vec{p}_1) + O_{i_1, i_2}(\vec{p}_1, \vec{p}_2) \right] F_{i_2}^{\vec{p}_2\,\dagger} F_{i_1}^{\vec{p}_1}$$

$$= \int d^3 p_1 d^3 p_2\, O_{1,1}(\vec{p}_2, \vec{p}_1) F_{i_2}^{\vec{p}_2\,\dagger} \varepsilon_{i_1 i_2} F_{i_1}^{\vec{p}_1} + H.c., \quad (3.29)$$

so that the expression for $\left[R^{(1)}, V\right]_{M_r, nucl}$ reaches the form:

$$\left[R^{(1)}, V\right]_{M_r, nucl} = -\int d^3 p_1 d^3 p_2 d^3 k d^3 q$$

$$\times F_{i_2}^{\vec{p}_2\,\dagger} \varepsilon_{i_1 i_2} F_{i_1}^{\vec{p}_1} \left[ R_{1,1}^{\vec{k}}(\vec{p}_2, \vec{q}) V_{1,1}^{-\vec{k}}(\vec{q}, \vec{p}_1) + R_{2,1}^{\vec{k}}(\vec{p}_2, \vec{q}) V_{1,2}^{-\vec{k}}(\vec{q}, \vec{p}_1) \right] + H.c.. \quad (3.30)$$

According to Eqs. (2.8) and (2.9), to fix the mass shifts we require:

$$\frac{1}{2}\left[R^{(1)}, V\right]_{M_r, g, mes} + M_{ren, g, mes} = 0, \quad (3.31)$$

$$\frac{1}{2}\left[R^{(1)}, V\right]_{M_r, g, nucl} + M_{ren, g, nucl} = 0, \quad (3.32)$$

after which it appears in our model that

$$M^{(2)}_{ren,b,rest}=0. \tag{3.33}$$

Using the conditions (3.31) and (3.32), we obtain the expressions for the shifts of particle masses:

$$\delta\mu^2(\vec{k})=4\omega_k\int d^3k_1 d^3p d^3q\left[V^{-\vec{k}}_{2,1}(\vec{q},\vec{p})R^{\vec{k}_1}_{1,2}(\vec{p},\vec{q})-R^{\vec{k}_1}_{2,1}(\vec{p},\vec{q})V^{-\vec{k}}_{1,2}(\vec{q},\vec{p})\right], \tag{3.34}$$

$$\delta m^2(\vec{p})=8E_p\int d^3p_1 d^3q d^3k\left[R^{\vec{k}}_{1,1}(\vec{p}_1,\vec{q})V^{-\vec{k}}_{1,1}(\vec{q},\vec{p})+R^{\vec{k}}_{2,1}(\vec{p}_1,\vec{q})V^{-\vec{k}}_{1,2}(\vec{q},\vec{p})\right]. \tag{3.35}$$

By means of simple algebraic transformations these expressions can be brought to the explicitly covariant forms, providing the momentum independence of the mass shifts and coincidence with the respective results derived in the Dyson-Feynman covariant formalism:

$$\delta\mu^2(k)=\delta\mu^2(\mu,0,0,0)=\frac{g^2}{(2\pi)^3}\int\frac{d^3p}{E_p}\left(\frac{\mu^2}{4(pk)^2-\mu^4}\right), \tag{3.36}$$

$$\delta m^2(p)=\delta m^2(m,0,0,0)=\frac{g^2}{(2\pi)^3}\int\frac{d^3k}{\omega_k}\frac{1}{-\mu^2+2pk}$$

$$+\frac{g^2}{(2\pi)^3}\int\frac{d^3k}{E_k}\frac{1}{\mu^2-2m^2+2pk}, \tag{3.37}$$

where $p=(E_p,\vec{p})$, $k=(\omega_k,\vec{k})$.

The expressions for the mass shifts in the field-theoretical model including the scalar mesons and the nucleons with spins were first obtained with help of the formalism of the clothing UT in Refs. [9,10].

### 3.3. Coupling constant renormalization

The charge shift can be obtained after collecting vertex-like operators in the third order in $g$ (2.16). The commutator $\left[R^{(1)}, V\right]^2$ in the model under study has the form:

$$\left[R^{(1)}, V\right]^2 = \int d^3k_1 d^3k_2 d^3k_3$$

$$\times \left[ a^\dagger_{\vec{k}_1} a^\dagger_{\vec{k}_2} a^\dagger_{\vec{k}_3} \hat{\Phi}_1\left(\vec{k}_3,\vec{k}_2,\vec{k}_1\right) + a^\dagger_{\vec{k}_3} a^\dagger_{\vec{k}_2} a_{\vec{k}_1} \hat{\Phi}_2\left(\vec{k}_3,\vec{k}_2,\vec{k}_1\right) \right.$$

$$\left. + a^\dagger_{\vec{k}_2} \hat{\Phi}_3\left(\vec{k}_1,\vec{k}_2\right) \delta\left(\vec{k}_1 - \vec{k}_3\right) \right] + H.c., \qquad (3.38)$$

where:

$$\hat{\Phi}_1\left(\vec{k}_3,\vec{k}_2,\vec{k}_1\right) = \left[\hat{R}^{\vec{k}_3},\left[\hat{R}^{\vec{k}_2},\hat{V}^{\vec{k}_1}\right]\right], \qquad (3.39)$$

$$\hat{\Phi}_2\left(\vec{k}_3,\vec{k}_2,\vec{k}_1\right) = \left[\hat{R}^{\vec{k}_3},\left[\hat{R}^{\vec{k}_2},\hat{V}^{-\vec{k}_1}\right]\right] - \left[\hat{R}^{\vec{k}_1\dagger},\left[\hat{R}^{\vec{k}_2},\hat{V}^{\vec{k}_3}\right]\right] - \left[\hat{R}^{\vec{k}_3\dagger},\left[\hat{R}^{\vec{k}_1},\hat{V}^{-\vec{k}_2}\right]\right],$$

$$(3.40)$$

$$\hat{\Phi}_3\left(\vec{k}_1,\vec{k}_2\right) = -2\hat{R}^{\vec{k}_1}\left[\hat{R}^{\vec{k}_2},V^{-\vec{k}_1}\right] - \left[\hat{R}^{\vec{k}_2},\hat{R}^{\vec{k}_1}\right]\hat{V}^{-\vec{k}_1} - 2\hat{R}^{\vec{k}_1\dagger}\left[\hat{R}^{\vec{k}_2},\hat{V}^{\vec{k}_1}\right]$$

$$-\hat{R}^{\vec{k}_1\dagger}\left[\hat{R}^{\vec{k}_1},\hat{V}^{\vec{k}_2}\right] - \left[\hat{R}^{\vec{k}_1},\hat{V}^{-\vec{k}_2}\right]^\dagger \hat{R}^{\vec{k}_1} - \left[\hat{R}^{\vec{k}_2},\hat{R}^{\dagger\vec{k}_1}\right]\hat{V}^{\vec{k}_1}. \qquad (3.41)$$

After normal ordering, the commutator $\left[R^{(1)}, V\right]^2$ can be represented as the following partition into the operators of different types:

$$\left[R^{(1)}, V\right]^2 = \left[R^{(1)}, V\right]^2_{V_r} + \left[R^{(1)}, V\right]^2_{t,b} + \left[R^{(1)}, V\right]^2_{t,g}.$$

The commutator $\left[R^{(1)}, V\right]^2_{t,g}$ includes all operators of the observable processes of the third order in the meson-nucleon system [4]:

$$\frac{1}{3}\left[R^{(1)},V\right]_{t,g}^{2} = H^{(3)}\left(\text{NN}\leftrightarrow\pi\text{NN}\right)+H^{(3)}\left(\text{N}\bar{\text{N}}\leftrightarrow\pi\text{N}\bar{\text{N}}\right)$$

$$+H^{(3)}\left(\bar{\text{N}}\bar{\text{N}}\leftrightarrow\pi\bar{\text{N}}\bar{\text{N}}\right)+H^{(3)}\left(\pi\text{N}\leftrightarrow\pi\pi\text{N}\right)+H^{(3)}\left(\pi\bar{\text{N}}\leftrightarrow\pi\pi\bar{\text{N}}\right)$$

$$+H^{(3)}\left(\text{N}\bar{\text{N}}\leftrightarrow\pi\pi\pi\right)+H^{(3)}\left(\pi\text{N}\bar{\text{N}}\leftrightarrow\pi\pi\right)$$

$$+H^{(3)}\left(\text{NN}\bar{\text{N}}\leftrightarrow\pi\text{N}\right)+H^{(3)}\left(\text{N}\bar{\text{N}}\bar{\text{N}}\leftrightarrow\pi\bar{\text{N}}\right). \tag{3.42}$$

For example, let us write down some of these operators:

$$H^{(3)}\left(\text{NN}\leftrightarrow\pi\text{NN}\right)=\int d^{3}p_{2}d^{3}q_{2}d^{3}p_{1}d^{3}q_{1}d^{3}k$$

$$\times V^{(\text{NN}\leftrightarrow\pi\text{NN})}\left(\vec{p}_{1},\vec{p}_{2},\vec{k};\vec{q}_{1},\vec{q}_{2}\right)F_{1}^{\vec{p}_{1}\dagger}F_{1}^{\vec{p}_{2}\dagger}F_{1}^{\vec{q}_{1}}F_{1}^{\vec{q}_{2}}a_{\vec{k}}^{\dagger}+H.c.,$$

$$V^{(\text{NN}\leftrightarrow\pi\text{NN})}\left(\vec{p}_{1},\vec{p}_{2},\vec{k};\vec{q}_{1},\vec{q}_{2}\right)=\frac{(-1)^{j+1}}{3}\int d^{3}k_{1}d^{3}q$$

$$\left\{-V_{1,1}^{\vec{k}_{1}}\left(\vec{p}_{2},\vec{q}_{2}\right)\cdot R_{1,j}^{\vec{k}}\left(\vec{p}_{1},\vec{q}\right)R_{1,j}^{\vec{k}_{1}}\left(\vec{q}_{1},\vec{q}\right)-2R_{1,1}^{\vec{k}_{1}}\left(\vec{p}_{2},\vec{q}_{2}\right)\cdot R_{1,j}^{\vec{k}}\left(\vec{p}_{1},\vec{q}\right)V_{j,1}^{-\vec{k}_{1}}\left(\vec{q},\vec{q}_{1}\right)\right.$$

$$-R_{1,1}^{\vec{k}_{1}}\left(\vec{p}_{2},\vec{q}_{2}\right)\cdot V_{1,j}^{\vec{k}}\left(\vec{p}_{1},\vec{q}\right)R_{1,j}^{\vec{k}_{1}}\left(\vec{q}_{1},\vec{q}\right)$$

$$-V_{1,1}^{-\vec{k}_{1}}\left(\vec{p}_{2},\vec{q}_{2}\right)\cdot R_{1,j}^{\vec{k}}\left(\vec{p}_{1},\vec{q}\right)R_{j,1}^{\vec{k}_{1}}\left(\vec{q},\vec{q}_{1}\right)-2R_{1,1}^{\vec{k}_{1}}\left(\vec{q}_{2},\vec{p}_{2}\right)\cdot R_{1,j}^{\vec{k}}\left(\vec{p}_{1},\vec{q}\right)V_{j,1}^{\vec{k}_{1}}\left(\vec{q},\vec{q}_{1}\right)$$

$$+R_{1,1}^{\vec{k}_{1}}\left(\vec{q}_{2},\vec{p}_{2}\right)\cdot V_{1,j}^{\vec{k}}\left(\vec{p}_{1},\vec{q}\right)R_{j,1}^{\vec{k}_{1}}\left(\vec{q},\vec{q}_{1}\right)$$

$$+V_{1,1}^{-\vec{k}_{1}}\left(\vec{p}_{2},\vec{q}_{2}\right)\cdot R_{1,j}^{\vec{k}_{1}}\left(\vec{p}_{1},\vec{q}\right)R_{j,1}^{\vec{k}}\left(\vec{q},\vec{q}_{1}\right)+2R_{1,1}^{\vec{k}_{1}}\left(\vec{p}_{2},\vec{q}_{2}\right)\cdot V_{1,j}^{-\vec{k}_{1}}\left(\vec{p}_{1},\vec{q}\right)R_{j,1}^{\vec{k}}\left(\vec{q},\vec{q}_{1}\right)$$

$$-R_{1,1}^{\vec{k}_{1}}\left(\vec{q}_{2},\vec{p}_{2}\right)\cdot R_{1,j}^{\vec{k}_{1}}\left(\vec{p}_{1},\vec{q}\right)V_{j,1}^{\vec{k}}\left(\vec{q},\vec{q}_{1}\right)$$

$$+V_{1,1}^{\vec{k}_{1}}\left(\vec{p}_{2},\vec{q}_{2}\right)\cdot R_{j,1}^{\vec{k}_{1}}\left(\vec{q},\vec{p}_{1}\right)R_{j,1}^{\vec{k}}\left(\vec{q},\vec{q}_{1}\right)+2R_{1,1}^{\vec{k}_{1}}\left(\vec{q}_{2},\vec{p}_{2}\right)\cdot V_{1,j}^{\vec{k}_{1}}\left(\vec{p}_{1},\vec{q}\right)R_{j,1}^{\vec{k}}\left(\vec{q},\vec{q}_{1}\right)$$

$$\left.+R_{1,1}^{\vec{k}_{1}}\left(\vec{p}_{2},\vec{q}_{2}\right)\cdot R_{j,1}^{\vec{k}_{1}}\left(\vec{q},\vec{p}_{1}\right)V_{j,1}^{\vec{k}}\left(\vec{q},\vec{q}_{1}\right)\right\}. \tag{3.43}$$

$$H^{(3)}\left(\text{N}\bar{\text{N}}\leftrightarrow\pi\pi\pi\right)=\int d^{3}p_{1}d^{3}p_{2}d^{3}k_{1}d^{3}k_{2}d^{3}k_{3}$$

$$\times V^{(\text{N}\bar{\text{N}}\leftrightarrow\pi\pi\pi)}\left(\vec{k}_{3},\vec{k}_{2},\vec{k}_{1};\vec{p}_{2},\vec{p}_{1}\right)F_{2}^{\vec{p}_{2}\dagger}F_{1}^{\vec{p}_{1}}a_{\vec{k}_{3}}^{\dagger}a_{\vec{k}_{2}}^{\dagger}a_{-\vec{k}_{1}}^{\dagger}+H.c.,$$

$$V^{(N\bar{N} \leftrightarrow \pi\pi\pi)}\left(\vec{k}_3,\vec{k}_2,\vec{k}_1;\vec{p}_2,\vec{p}_1\right) = \frac{(-1)^{j+i}}{3}\int d^3p\, d^3q$$
$$\times \left\{ R_{2,i}^{\vec{k}_3}(\vec{p}_2,\vec{p}) R_{i,j}^{\vec{k}_2}(\vec{p},\vec{q}) V_{j,1}^{-\vec{k}_1}(\vec{q},\vec{p}_1) - 2 R_{2,j}^{\vec{k}_3}(\vec{p}_2,\vec{q}) V_{j,i}^{-\vec{k}_1}(\vec{q},\vec{p}) R_{i,1}^{\vec{k}_2}(\vec{p},\vec{p}_1) \right.$$
$$\left. + V_{2,i}^{-\vec{k}_1}(\vec{p}_2,\vec{p}) R_{i,j}^{\vec{k}_2}(\vec{p},\vec{q}) R_{j,1}^{\vec{k}_3}(\vec{q},\vec{p}_1) \right\}. \tag{3.44}$$

Operators contained in $\left[R^{(1)},V\right]^2_{t,b}$ (e.g., $b^\dagger b a^\dagger a^\dagger$, $bbdda^\dagger$) must be eliminated from the Hamiltonian by the third clothing UT.

Besides the bad transition operators of the third order, the Hamiltonian contains the vertex-like operators of the third order which, after collecting and fixing the value of the coupling constant shift, have to be eliminated. After the normal ordering of the nucleonic operators in $a^\dagger_{\vec{k}_3} \hat{\Phi}_3^{\vec{k}_1,\vec{k}_3}$, we separate the vertex-like operators of the $g^3$-order:

$$\left[R^{(1)},V\right]^2_{V_r} = \left[R^{(1)},V\right]^2_{V_r}(ChargeRen)$$
$$+ \left[R^{(1)},V\right]^2_{V_r}(LegRenMes) + \left[R^{(1)},V\right]^2_{V_r}(LegRenNucl). \tag{3.45}$$

In this sum the first item corresponds to the vertex clothing (charge renormalization) in the third order, while the second and third ones are responsible for the renormalization of the meson and nucleon wave functions (leg renormalization). Commutator $\left[R^{(1)},M^{(2)}_{ren}\right]$ in the considered model contains only the vertex-like operators corresponding to the wave function renormalization. As we will show, all of these operators (with the account for the results of mass renormalization in the second order (3.34), (3.35)) are cancelled via the vertex-like operators in $\left[R^{(1)},V\right]^2_{V_r}(LegRenMes)$ and $\left[R^{(1)},V\right]^2_{V_r}(LegRenNucl)$.

Operators responsible for the renormalization of the meson wave function have the form:

$$\left[R^{(1)},V\right]^2_{V_r}(LegRenMes)=\int d^3k\,d^3p\,d^3q$$
$$\times\left[3I^{(1)}_{i,j}(\vec{p},\vec{q},\vec{k})+I^{(2)}_{i,j}(\vec{p},\vec{q},\vec{k})\right]F^{i\dagger}_{\vec{p}}F^{j}_{\vec{q}}a^\dagger_{\vec{k}}+H.c., \qquad (3.46)$$

$$I^{(1)}_{i,j}(\vec{p},\vec{q},\vec{k})=\int d^3p_1\,d^3q_1\,d^3k_1$$
$$\times\left[R^{\vec{k}_1}_{i,j}(\vec{p},\vec{q})+R^{-\vec{k}_1}_{j,i}(\vec{q},\vec{p})\right]\left[V^{-\vec{k}_1}_{-1,1}(\vec{q}_1,\vec{p}_1)R^{\vec{k}}_{1,-1}(\vec{p}_1,\vec{q}_1)-R^{\vec{k}}_{-1,1}(\vec{q}_1,\vec{p}_1)V^{\vec{k}_1}_{1,-1}(\vec{p}_1,\vec{q}_1)\right]$$
$$=\frac{1}{4}\int d^3q_1\left[R^{\vec{k}}_{i,j}(\vec{p},\vec{q})+R^{-\vec{k}}_{j,i}(\vec{q},\vec{p})\right]\omega_{q_1}\delta\mu^2(q_1), \qquad (3.47)$$

$$I^{(2)}_{i,j}(\vec{p},\vec{q},\vec{k})=\int d^3p_1\,d^3q_1\,d^3k_1$$
$$\times V^{\vec{k}_1}_{i,j}(\vec{p},\vec{q})\left[R^{\vec{k}_1}_{1,-1}(\vec{p}_1,\vec{q}_1)R^{\vec{k}}_{1,-1}(\vec{p}_1,\vec{q}_1)-R^{\vec{k}}_{-1,1}(\vec{p}_1,\vec{q}_1)R^{\vec{k}_1}_{-1,1}(\vec{p}_1,\vec{q}_1)\right]. \qquad (3.48)$$

Having derived the commutator

$$\left[R^{(1)},M^{(2)}_{ren,mes}\right]=-\frac{1}{4}\int d^3k_1\,d^3p_1\,d^3q_1\,d^3k$$
$$\times\left[R^{\vec{k}}_{i_1,j_1}(\vec{p}_1,\vec{q}_1)+R^{-\vec{k}}_{j_1,i_1}(\vec{q}_1,\vec{p}_1)\right]\omega_{\vec{k}_1}\delta\mu^2(\vec{k}_1)F^{\vec{p}_1\dagger}_{i_1}F^{\vec{q}_1}_{j_1}a^\dagger_{\vec{k}}+H.c., \qquad (3.49)$$

we see that it is totally cancelled by the first part (3.47) of the operator $\left[R^{(1)},V\right]^2_{V_r}(LegRenMes)$. The remaining terms (3.48) enter the operator $V^{(3)}_{ren,rest}$ (2.16) and will be eliminated from the Hamiltonian via the third clothing UT.

Operators responsible for the renormalization of the nucleon wave function are as follows:

$$\left[R^{(1)},V\right]^2 (LegRenNucl) = \int d^3p\, d^3q\, d^3k$$
$$\times \left[3O^{(1)}_{i,j}(\vec{p},\vec{q},\vec{k}) + O^{(2)}_{i,j}(\vec{p},\vec{q},\vec{k})\right] F^{\vec{p}\dagger}_i F^{\vec{q}}_j a^{\dagger}_{\vec{k}} + H.c., \quad (3.50)$$

$$O^{(1)}_{i,j}(\vec{p},\vec{q},\vec{k}) = (-1)^{i_1+1}\varepsilon_{ij} \int d^3k_1\, d^3p_1\, d^3q_1$$
$$\times \left[R^{\vec{p},\vec{p}_1}_{1,i_1}(\vec{k}) + R^{\vec{p}_1,\vec{p}}_{i_1,1}(\vec{k})\right]\left[R^{\vec{q}_1,\vec{q}}_{2,1}(\vec{k}_1) V^{\vec{p}_1,\vec{q}_1}_{1,2}(\vec{k}_1) + R^{\vec{q},\vec{q}_1}_{1,1}(\vec{k}_1) V^{\vec{q}_1,\vec{p}_1}_{1,1}(\vec{k}_1)\right]$$
$$= (-1)^{i_1+1}\varepsilon_{i,j}\frac{1}{8}\int d^3k_1\left[R^{\vec{p},\vec{q}}_{1,i_1}(\vec{k}) + R^{\vec{q},\vec{p}}_{i_1,1}(\vec{k})\right] E_{k_1}\delta m^2(k_1), \quad (3.51)$$

$$O^{(2)}_{i,j}(\vec{p},\vec{q},\vec{k}) = (-1)^{i_1+1}\int d^3k_1\, d^3p_1\, d^3q_1$$
$$\times \Big\{\left[R^{\vec{p},\vec{p}_1}_{i,2}(\vec{k}_1) R^{\vec{q}_1,\vec{p}_1}_{i_1,2}(\vec{k}_1) - R^{\vec{p}_1,\vec{p}}_{1,i}(\vec{k}_1) R^{\vec{p}_1,\vec{q}_1}_{1,i_1}(\vec{k}_1)\right] V^{\vec{q}_1,\vec{q}}_{i_1,j}(\vec{k})$$
$$+ V^{\vec{p},\vec{q}_1}_{i,i_1}(\vec{k})\left[R^{\vec{q}_1,\vec{p}_1}_{i_1,2}(\vec{k}_1) R^{\vec{q},\vec{p}_1}_{j,2}(\vec{k}_1) - R^{\vec{p}_1,\vec{q}_1}_{1,i_1}(\vec{k}_1) R^{\vec{p}_1,\vec{q}}_{1,j}(\vec{k}_1)\right]\Big\}. \quad (3.52)$$

Having derived the commutator

$$\left[R^{(1)}, M^{(2)}_{ren,nucl}\right] = \frac{1}{8}(-1)^{i_1}\int d^3k_1\, d^3p\, d^3p\, d^3k$$
$$\times \left[R^{\vec{p},\vec{q}}_{1,i_1}(\vec{k}) + R^{\vec{q},\vec{p}}_{i_1,1}(\vec{k})\right] E_{k_1}\delta m^2(\vec{k}_1) F^{\vec{p}\dagger}_i \varepsilon_{ij} F^{\vec{q}}_j a^{\dagger}_{\vec{k}} + H.c., \quad (3.53)$$

we see that it is totally canceled with the respective operators contained in $\left[R^{(1)},V\right]^2 (LegRenNucl)$. Remaining operators in $\left[R^{(1)},V\right]^2 (LegRenNucl)$ enter the operator $V^{(3)}_{ren,rest}$ (2.16).

Let us extract the operator $\left[R^{(1)},V\right]^2_{V_r}(ChargeRen)$ which is responsible for the charge renormalization from the commutator $\left[R^{(1)},V\right]^2$:

$$\left[R^{(1)},V\right]^2_{V_r}(ChargeRen) = -\int d^3p\, d^3q\, d^3k$$

$$\times J_{i,j}\left(\vec{p},\vec{q},\vec{k}\right)F_i^{\vec{p}\dagger}F_j^{\vec{q}}a_{\vec{k}}^{\dagger} + H.c., \tag{3.54}$$

$$J_{i,j}\left(\vec{p},\vec{q},\vec{k}\right) = \int d^3p_1\, d^3q_1\, d^3k_1$$

$$\times \Big[ -R_{i,1}^{\vec{p},\vec{p}_1}\left(\vec{k}_1\right)R_{1,1}^{\vec{p}_1,\vec{q}_1}\left(\vec{k}\right)V_{1,j}^{\vec{q}_1,\vec{q}}\left(-\vec{k}_1\right) + 2R_{i,1}^{\vec{p},\vec{p}_1}\left(\vec{k}_1\right)R_{1,2}^{\vec{p}_1,\vec{q}_1}\left(\vec{k}\right)V_{2,j}^{\vec{q}_1,\vec{q}}\left(-\vec{k}_1\right)$$

$$-R_{i,2}^{\vec{p},\vec{p}_1}\left(\vec{k}_1\right)R_{2,1}^{\vec{p}_1,\vec{q}_1}\left(\vec{k}\right)V_{1,j}^{\vec{q}_1,\vec{q}}\left(-\vec{k}_1\right) - R_{1,i}^{\vec{p}_1,\vec{p}}\left(\vec{k}_1\right)R_{1,1}^{\vec{p}_1,\vec{q}_1}\left(\vec{k}\right)V_{1,j}^{\vec{q}_1,\vec{q}}\left(\vec{k}_1\right)$$

$$+2R_{1,i}^{\vec{p}_1,\vec{p}}\left(\vec{k}_1\right)R_{1,2}^{\vec{p}_1,\vec{q}_1}\left(\vec{k}\right)V_{2,j}^{\vec{q}_1,\vec{q}}\left(\vec{k}_1\right) - R_{2,i}^{\vec{p}_1,\vec{p}}\left(\vec{k}_1\right)R_{2,1}^{\vec{p}_1,\vec{q}_1}\left(\vec{k}\right)V_{1,j}^{\vec{q}_1,\vec{q}}\left(\vec{k}_1\right)$$

$$-V_{i,2}^{\vec{p},\vec{p}_1}\left(-\vec{k}_1\right)R_{2,2}^{\vec{p}_1,\vec{q}_1}\left(\vec{k}\right)R_{2,j}^{\vec{q}_1,\vec{q}}\left(\vec{k}_1\right) + 2V_{i,1}^{\vec{p},\vec{p}_1}\left(-\vec{k}_1\right)R_{1,2}^{\vec{p}_1,\vec{q}_1}\left(\vec{k}\right)R_{2,j}^{\vec{q}_1,\vec{q}}\left(\vec{k}_1\right)$$

$$-V_{i,2}^{\vec{p},\vec{p}_1}\left(-\vec{k}_1\right)R_{2,1}^{\vec{p}_1,\vec{q}_1}\left(\vec{k}\right)R_{1,j}^{\vec{q}_1,\vec{q}}\left(\vec{k}_1\right) - V_{i,2}^{\vec{p},\vec{p}_1}\left(-\vec{k}_1\right)R_{2,2}^{\vec{p}_1,\vec{q}_1}\left(\vec{k}\right)R_{j,2}^{\vec{q},\vec{q}_1}\left(\vec{k}_1\right)$$

$$+2V_{i,1}^{\vec{p},\vec{p}_1}\left(\vec{k}_1\right)R_{1,2}^{\vec{p}_1,\vec{q}_1}\left(\vec{k}\right)R_{j,2}^{\vec{q},\vec{q}_1}\left(\vec{k}_1\right) - V_{i,2}^{\vec{p},\vec{p}_1}\left(-\vec{k}_1\right)R_{2,1}^{\vec{p}_1,\vec{q}_1}\left(\vec{k}\right)R_{j,1}^{\vec{q},\vec{q}_1}\left(\vec{k}_1\right)$$

$$-R_{i,1}^{\vec{p},\vec{p}_1}\left(\vec{k}_1\right)V_{1,2}^{\vec{p}_1,\vec{q}_1}\left(\vec{k}\right)R_{j,2}^{\vec{q},\vec{q}_1}\left(\vec{k}_1\right) + 2R_{i,2}^{\vec{p},\vec{p}_1}\left(\vec{k}_1\right)V_{2,2}^{\vec{p}_1,\vec{q}_1}\left(\vec{k}\right)R_{j,2}^{\vec{q},\vec{q}_1}\left(\vec{k}_1\right)$$

$$-R_{1,i}^{\vec{p}_1,\vec{p}}\left(\vec{k}_1\right)V_{1,2}^{\vec{p}_1,\vec{q}_1}\left(\vec{k}\right)R_{2,j}^{\vec{q}_1,\vec{q}}\left(\vec{k}_1\right) + 2R_{1,i}^{\vec{p}_1,\vec{p}}\left(\vec{k}_1\right)V_{1,1}^{\vec{p}_1,\vec{q}_1}\left(\vec{k}\right)R_{1,j}^{\vec{q}_1,\vec{q}}\left(\vec{k}_1\right)$$

$$-R_{2,i}^{\vec{p}_1,\vec{p}}\left(\vec{k}_1\right)V_{2,1}^{\vec{p}_1,\vec{q}_1}\left(\vec{k}\right)R_{1,j}^{\vec{q}_1,\vec{q}}\left(\vec{k}_1\right) \Big] + H.c.. \tag{3.55}$$

The operator structure of the expression (3.54) repeats the structure of the vertex counterterm $V_{ren}$ but the kernels of these operators are different. Comparing the matrix elements of the kernel of operator (3.54) corresponding to the "diagonal" structures $b^{\dagger}ba^{\dagger}$ and $dd^{\dagger}a^{\dagger}$, we find them being equal, manifestly confirming the CPT invariance of the model at hands. Thus, to fix the charge correction we cancel the "diagonal" parts of the operators $V_{ren}$ (3.13) and $\left[R^{(1)},V\right]^2_{V_r}(ChargeRen)$ (3.54). As the "non-diagonal" parts of these operators are bad anyway, they will enter the operator $V^{(3)}_{ren,rest}$ collecting bad terms of the third order which must be removed via the third clothing UT. This way we get

$$\delta g^{(3)} = \frac{g^3}{8(2\pi)^{9/2}} \int \frac{d^3 k'}{E_{p-k'} E_{q-k'} \omega_{k'}}$$

$$\times \left\{ \Delta \left( D^{1,2}_{q-k',q,k'}, D^{1,2}_{p-k',q-k',k}, -D^{1,2}_{p-k',p,k'} \right) + \Delta \left( -D^{1,2}_{q-k',q,k'}, D^{2,2}_{p-k',p,k'}, -D^{2,2}_{p-k',q-k',k} \right) \right.$$

$$+ \Delta \left( -D^{1,1}_{q-k',q,k'}, D^{2,1}_{p-k',q-k',k}, D^{1,1}_{p-k',p,k'} \right) + \Delta \left( -D^{1,2}_{p-k',p,k'}, D^{2,2}_{q-k',q,k'}, D^{1,1}_{p-k',q-k',k} \right)$$

$$\left. + \Delta \left( -D^{1,1}_{p-k',p,k'}, D^{2,1}_{q-k',q,k'}, -D^{2,2}_{p-k',q-k',k} \right) + \Delta \left( -D^{1,1}_{q-k',q,k'}, D^{2,1}_{p-k',p,k'}, D^{1,1}_{p-k',q-k',k} \right) \right\}, (3.56)$$

where we adopt the following denotations : $\Delta(a,b,c) = (ab + bc - 2ac)/3$ and $D^{i,j}_{p,q,k} = \left[ (-1)^{i+1} E_p + (-1)^{j+1} E_q + \omega_k \right]^{-1}$, $i, j = 1,2$. Expression (3.56) has the 3-momentum shell $\vec{q} = \vec{p} + \vec{k}$.

3.4 Mechanisms of the charge renormalization

Each of the six items in (3.56) exposes one of the six mechanisms via which the charge renormalization holds in the third order in the coupling constant (Fig. 1). Namely, the first item in (3.56) corresponds to the diagram $a$ in Fig. 1, the second item corresponds to the diagram $b$ and so on. The direction of arrows in Fig. 1 differs particles from antiparticles and has nothing to do with the chronology of events because we work in the Schrödinger picture.

These diagrams are similar to those used in the old-fashioned perturbation theory and their topology is conditioned only by the 3-momentum conservation in each vertex. However, it appears that each such graph represents three algebraic structures comprising the expression (3.56). This means that the diagrammatic language of the old-fashioned perturbation theory is too poor for representing all the algebraic details of our perturbation approach.

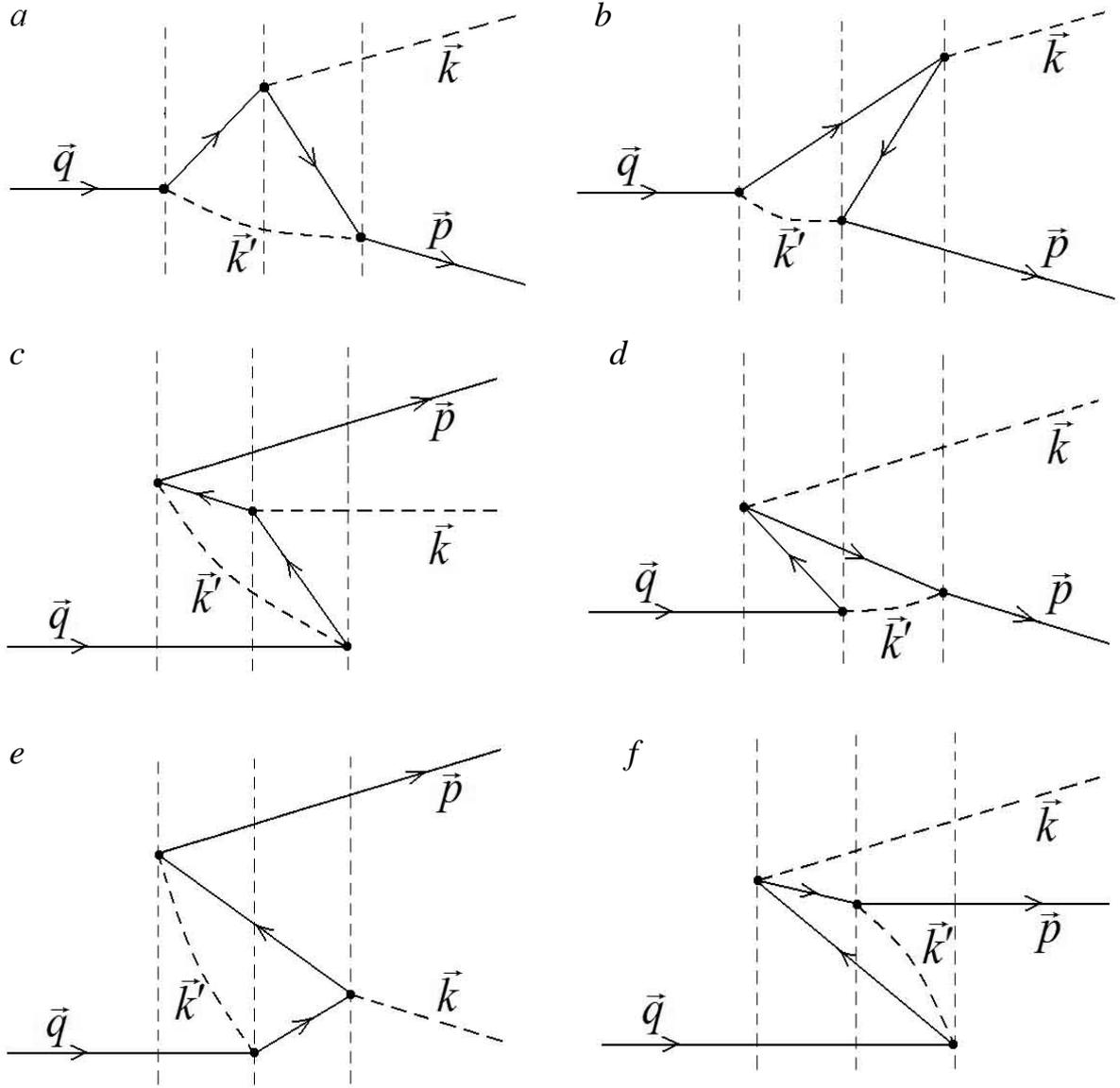

Fig. 1. Graphical representation of the expression (3.56). Diagram *a* corresponds to the first term in (3.56), diagram *b* corresponds to the second one and so on. Arrows differ particles from antiparticles.

Being concerned much in having more deep penetration into the problem of formation of the off-energy-shell structures in our perturbation approach, here we are focused on finding an adequate graphical technique to represent our algebra. Searching for the support in this question, we notice that each of the non-covariant three-energy propagators $D^{i,j}_{p,q,k}$ entering the expression for the charge shift corresponds to that particular interaction vertex where the energy conser-vation cannot hold (the propagator $D^{i,j}_{p,q,k}$ has a pole). So, it is instructive to take the graph typical of the old-fashioned perturbation theory and directly mark by open circles the vertices corresponding to the three-energy propagators $D^{i,j}_{p,q,k}$.

Following this guideline, we take each of the graphs shown in Fig. 1 and decompose it into the three novel diagrams according to the assumption that the energy conservation does not hold in two from three vertices appearing in the third order in the coupling constant. As an example, in Fig. 2 we show such a decomposition for the graph *a* from Fig. 1 and the respective algebraic structures.

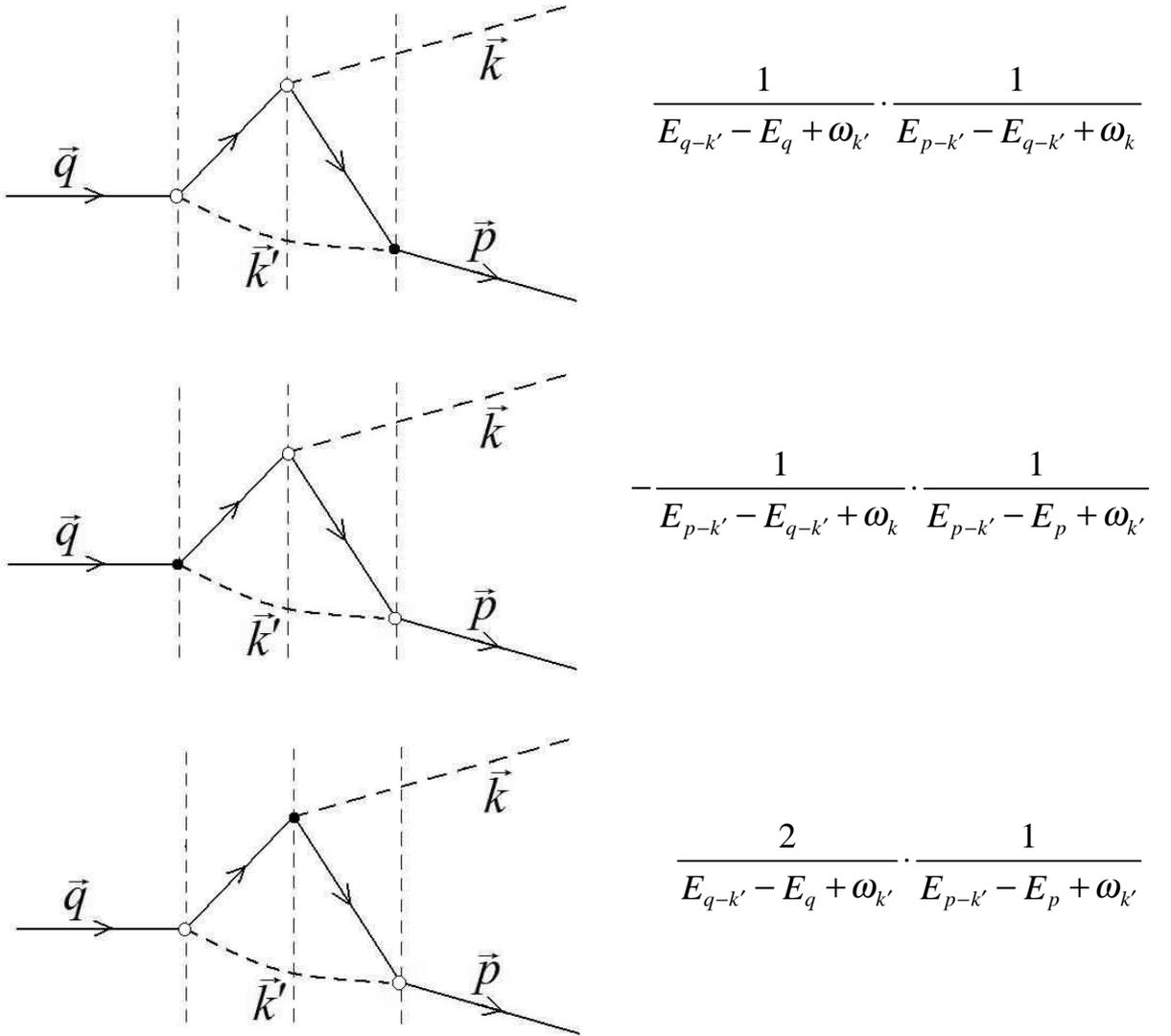

Fig. 2. Diagram *a* from Fig. 1 decomposed into three novel graphs (on the left) and the corresponding algebraic structures from Eq. (3.56).

To proceed further we make use of the fact that the amount of the old-fashioned diagrams, which always represent the on-energy-shell contributions, appears not enough to provide the one-to-one correspondence for our non-covariant algebra. This means that the derived charge correction in the form (3.56) is the function off the energy shell and we need to specify that part of Eq. (3.56) which contains the actual value for the charge correction in question. To do this we simply put Eq. (3.56) on the energy shell, looking for the following decomposition:

$$\delta g^{(3)} = \delta g^{(3)}_{on-energy-shell} + \delta g^{(3)}_{off-energy-shell}, \qquad (3.57)$$

where $\delta g^{(3)}_{off-energy-shell}$ goes to zero on the energy shell (i.e., if the energy conservation is implied), while $\delta g^{(3)}_{on-energy-shell}$ gives us the desired result for the charge correction. As is shown in the forthcoming section, the latter can be brought to the explicitly covariant form that depends only on the Lorenz-scalar combinations built of the particle momenta. After that the remaining part $\delta g^{(3)}_{off-energy-shell}$ enters the operator $V^{(3)}_{ren,rest}$ and will be eliminated from the Hamiltonian.

3.5. Reduction of the expression for the charge shift to the explicitly covariant form

To find the contribution $\delta g^{(3)}_{on-energy-shell}$ to the Eq. (3.56) we put the latter on the energy shell $E_p = E_q - \omega_k$ [14-16] and providing some transparent substitutions of variables get:

$$\delta g^{(3)} = \frac{g^3}{8(2\pi)^{9/2}} \int \frac{d^3 k'}{E_p E_q \omega_{k'}} (I_1 + I_2 + I_3 + I_4 + I_5 + I_6), \qquad (3.58)$$

where

$$I_1 = -D^{2,2}_{q,p,k}D^{2,1}_{q-k',q,k'}, \quad I_2 = D^{2,1,1}_{q-k',p,k,k'}D^{1,1}_{q,p,k}, \quad I_3 = D^{2,1,1}_{q-k',p,k,k'}D^{2,1}_{q-k',q,k'},$$

$$I_4 = D^{1,1}_{q,p,k}D^{1,1}_{q-k',q,k'}, \quad I_5 = -D^{1,1,2}_{q-k',p,k,k'}D^{2,2}_{q,p,k}, \quad I_6 = D^{1,1,2}_{q-k',p,k,k'}D^{1,1}_{q-k',q,k'}, \quad (3.59)$$

$\vec{q} = \vec{p} + \vec{k}$ and the four-energy propagators are denoted as:

$$D^{i,j,l}_{p,q,k,k'} \equiv \frac{1}{(-1)^{i+1}E_p + (-1)^{j+1}E_q + (-1)^{l+1}\omega_k + \omega_{k'}}, \quad i, j, l = 1, 2. \quad (3.60)$$

In striving to form the covariant structures from the three- and four-energy non-covariant propagators, we turn to the following algebraic trick. First, cut several propagators from (3.59) as follows:

$$I_3 \equiv I_7 + I_8,$$
$$I_7 = D^{2,1}_{q-k',q,k'}D^{2,1}_{q,p,k}, \qquad I_8 = -D^{2,1,1}_{q-k',p,k,k'}D^{2,1}_{q,p,k}, \qquad (3.61)$$

$$I_4 \equiv I_{11} + I_{12},$$
$$I_9 = D^{1,1}_{q,p,k}D^{1,2,2}_{q-k',p,k,k'}, \qquad I_{10} = -D^{1,1}_{q-k',q,k'}D^{1,2,2}_{q-k,p,k,k'}, \qquad (3.62)$$

$$I_5 \equiv I_9 + I_{10},$$
$$I_{11} = -D^{2,2}_{q,p,k}D^{1,2}_{q-k',q,k'}, \qquad I_{12} = -D^{1,1,2}_{q-k',p,k,k'}D^{1,2}_{q-k',q,k'}. \qquad (3.63)$$

Then, perform the following assembling (with addition and simultaneous subtraction of the same term):

$$I_1 + I_7 + I_{11} + (I_{13} - I_{13}) =$$
$$= \left(D^{2,1}_{q-k',q,k'} + D^{1,2}_{q-k',q,k'}\right)\left(D^{2,1}_{q,p,k} - D^{2,2}_{q,p,k}\right) - I_{13}, \qquad I_{13} = D^{1,2}_{q-k',q,k'}D^{2,1}_{q,p,k}, \qquad (3.64)$$

$$I_2 + I_8 + I_9 - (I_{14} - I_{14}) =$$

$$= \left(D^{1,1}_{q,p,k} - D^{2,1}_{q,p,k}\right)\left(D^{2,1,1}_{q-k',p,k,k'} + D^{1,2,2}_{q-k',p,k,k'}\right) - I_{14}, \qquad I_{14} = -D^{1,2,2}_{q-k',p,k,k'} D^{2,1}_{q,p,k}, \quad (3.65)$$

$$I_6 + I_{10} + I_{12} + (I_{15} - I_{15}) =$$

$$= \left(D^{1,1}_{q-k',q,k'} - D^{1,2}_{q-k',q,k'}\right)\left(D^{1,1,2}_{q-k',p,k,k'} - D^{1,2,2}_{q-k',p,k,k'}\right) - I_{15}, \quad I_{15} = D^{1,2,2}_{q-k,p,k,k'} D^{1,2}_{q-k',q,k'}. \quad (3.66)$$

After all that, it only remains to see that the sum

$$I_{13} + I_{14} + I_{15} = 0, \qquad (3.67)$$

to find the following expression for the charge correction $\delta g^{(3)}_{on-energy-shell}$:

$$\delta g^{(3)}_{on-energy-shell} = \frac{g^3}{8(2\pi)^3} \int \frac{d^3k'}{E_{p-k'} E_{q-k'} \omega_{k'}}$$

$$\times \left[ \frac{2\omega_{k'}}{(E_{q-k'} - E_q + \omega_{k'})(-E_{q-k'} + E_q + \omega_{k'})} \cdot \frac{-2E_{q-k'}}{(E_{p-k'} - E_{q-k'} + \omega_k)(-E_{q-k'} - E_{p-k'} + \omega_k)} \right.$$

$$+ \frac{-2E_{q-k'}}{(E_{q-k'} + E_{p-k'} + \omega_k)(-E_{q-k'} + E_{p-k'} + \omega_k)} \cdot \frac{2\omega_{k'}}{(-E_p + E_{p-k'} + \omega_{k'})(E_p - E_{p-k'} + \omega_{k'})}$$

$$\left. + \frac{-2E_{q-k'}}{(E_{q-k'} + E_q + \omega_{k'})(E_q - E_{q-k'} + \omega_{k'})} \cdot \frac{-2E_{p-k'}}{(E_p + E_{p-k'} + \omega_{k'})(E_p - E_{p-k'} + \omega_{k'})} \right], (3.68)$$

or in other form

$$\delta g^{(3)}_{on-energy-shell} = \frac{g^3}{8(2\pi)^3} \int \frac{d^3k'}{E_{p-k'} E_{q-k'} \omega_{k'}}$$

$$\times \left[ \frac{2\omega_{k'}}{\left(E_{q-k'}-E_q\right)^2 - \omega_{k'}^2} \cdot \frac{2E_{p-k'}}{\left(E_{q-k'}-\omega_k\right)^2 - E_{p-k'}^2} \right.$$

$$+ \frac{2E_{q-k'}}{\left(E_{p-k'}+\omega_k\right)^2 - E_{q-k'}^2} \cdot \frac{2\omega_{k'}}{\left(E_p - E_{p-k'}\right)^2 - \omega_{k'}^2}$$

$$\left. + \frac{2E_{q-k'}}{\left(E_q+\omega_{k'}\right)^2 - E_{q-k'}^2} \cdot \frac{2E_{p-k'}}{\left(E_p+\omega_{k'}\right)^2 - E_{p-k'}^2} \right], \tag{3.69}$$

determined on the 4-momentum shell $q = p + k$. Each of the items in the expression (3.69) can be presented in the explicitly covariant form via implementing several simple algebraic transformations. Really, let us consider the first term in Eq. (3.69):

$$\int \frac{d^3k'}{E_{p-k'}E_{q-k'}\omega_{k'}} \frac{2\omega_{k'}}{\left(E_{q-k'}-E_q\right)^2 - \omega_{k'}^2} \cdot \frac{2E_{p-k'}}{\left(E_{q-k'}-\omega_k\right)^2 - E_{p-k'}^2}$$

$$= -4\int \frac{d^3k'}{E_{q-k'}} \cdot \frac{1}{\mu^2 - 2m^2 + 2E_{q-k'}E_q - 2(\vec{q}-\vec{k'})\vec{q}} \cdot \frac{1}{\mu^2 - 2E_{q-k'}\omega_k + 2(\vec{q}-\vec{k'})\vec{k}}. \tag{3.70}$$

Providing the substitution of the integration variable $\vec{p} = \vec{q} - \vec{k'}$ and introducing the 4-momenta $q = (E_q, \vec{q})$, $k = (\omega_k, \vec{k})$, $p = (E_p, \vec{p})$, we set the equation:

$$\int \frac{d^3k'}{E_{p-k'}E_{q-k'}\omega_{k'}} \frac{2\omega_{k'}}{\left(E_{q-k'}-E_q\right)^2 - \omega_{k'}^2} \cdot \frac{2E_{p-k'}}{\left(E_{q-k'}-\omega_k\right)^2 - E_{p-k'}^2}$$

$$= -4\int \frac{d^3p}{E_p} \frac{1}{\mu^2 - 2m^2 + 2pq} \cdot \frac{1}{\mu^2 - 2pk}, \tag{3.71}$$

where $pq = E_p E_q - \vec{p}\vec{q}$ and $pk = E_p \omega_k - \vec{p}\vec{k}$. Applying such a recipe for other two items in (3.69), we find the following explicitly covariant expression for the charge correction [11-13]:

$$\delta g^{(3)}_{on-energy-shell} = -\frac{1}{2}\frac{g^3}{(2\pi)^3}\left[\int\frac{d^3p'}{E_{p'}}\frac{1}{(\mu^2 - 2p'k)(\mu^2 - 2m^2 - 2p'p)} + \right.$$

$$\left. -\int\frac{d^3k'}{\omega_{k'}}\frac{1}{(\mu^2 + 2k'p)(\mu^2 + 2k'q)} + \int\frac{d^3p'}{E_{p'}}\frac{1}{(\mu^2 + 2p'k)(\mu^2 - 2m^2 + 2p'q)}\right., \quad (3.72)$$

where $p' = (E_{p'}, \vec{p}')$, $k' = (\omega_{k'}, \vec{k}')$. The respective result for the charge correction in the model of nucleons with spins has been derived in [14-16].

3.6. Clothed particle representation vs. old-fashioned perturbation theory

In this section we are going to establish some general links between the perturbation series appearing in the unitary clothing transformation approach and the series typical of the old-fashioned perturbation theory (OFPT). Here we should bear in mind that the OFPT, being explicitly not covariant, is in fact equivalent to the purely covariant time-dependent Dyson-Feynman perturbation theory. So, if we succeed in striving to find certain common ground for both the clothing approach and the OFPT we immediately get the proof of the covariance and, therefore, the momentum independence of the charge shift derived in this work. The OFPT originates from the Lippmann-Schwinger equation for the $t$-matrix in the form [26]:

$$\langle f|t|i\rangle = \langle f|V|i\rangle + \lim_{\varepsilon \to +0}\int d^3k \frac{\langle f|V|k\rangle\langle k|t|i\rangle}{E_i - E_k + i\varepsilon}, \quad (3.73)$$

where $|i\rangle$, $|f\rangle$ and $|k\rangle$ are the eigenstates of the free Hamiltonian $H_F$ with the energies $E_i$, $E_f$ and $E_k$, respectively; $V$ is the interaction operator in the decomposition $H = H_F + V$. Symbol $\int d^3k$ denotes integration and summation over all possible momenta and spin variables (if any) of the particles in the intermediate state $|k\rangle$. Iterations of the Eq. (3.73) give the perturbative series for the *t*-matrix:

$$\langle f|t|i\rangle = \langle f|V|i\rangle + \lim_{\varepsilon \to +0} \int d^3k \frac{\langle f|V|k\rangle\langle k|V|i\rangle}{E_i - E_k + i\varepsilon}$$
$$+ \lim_{\substack{\varepsilon_1 \to +0 \\ \varepsilon_2 \to +0}} \int d^3k_1 d^3k_2 \frac{\langle f|V|k_1\rangle\langle k_1|V|k_2\rangle\langle k_2|V|i\rangle}{\left(E_i - E_{k_1} + i\varepsilon_1\right)\left(E_i - E_{k_2} + i\varepsilon_2\right)} + \ldots . \quad (3.74)$$

The series in the multiple commutators that appear in the Hamiltonian after the first clothing transformation, sandwiched between the same initial and final states, look as follows (to set link with Eq. (3.74) we simply omit respective counterterms):

$$\langle f|H|i\rangle = \sum_{n=1}^{\infty}\langle f|H^{(n)}|i\rangle = \langle f|V|i\rangle + \sum_{n=2}^{\infty}\frac{n-1}{n!}\langle f|[R,V]^{n-1}|i\rangle$$
$$= \langle f|V|i\rangle + \sum_{n=2}^{\infty}\frac{n-1}{n!}\sum_{k=0}^{n-1}(-1)^k C_{n-1}^k \langle f|R^{n-k-1}VR^k|i\rangle. \quad (3.75)$$

Further we need to compare the terms of the equal orders in the interaction operator $V$ in the r.h.s. of Eqs. (3.74) and (3.75). The first terms obviously coincide, so we turn to the second ones. Using the integral representation for the generator $R$ of the first clothing transformation (see [4,9]):

$$R = -i \lim_{\varepsilon \to +0} \int_0^{\infty} dt\, e^{-\varepsilon t} V(t), \quad (3.76)$$

where $V(t) = e^{iH_F t} V e^{-iH_F t}$ is the interaction operator in the Dirac picture, we find in the second order in $V$:

$$\langle f|H^{(2)}|i\rangle = \frac{1}{2}\langle f|RV - VR|i\rangle = -\frac{i}{2}\lim_{\varepsilon \to +0}\int_0^\infty dt\, e^{-\varepsilon t}\langle f|V(t)V - VV(t)|i\rangle. \quad (3.77)$$

Bearing in mind that all states we operate with are, by definition, the eigenstates of $H_F$, e.g., $\langle f|e^{iH_F t} = \langle f|e^{iE_f t}$ and $e^{iH_F t}|i\rangle = e^{iE_i t}|i\rangle$, and applying the unity operator $\int d^3k |k\rangle\langle k|$ with the full set of states, we find:

$$\langle f|H^{(2)}|i\rangle = -\frac{i}{2}\int d^3k\, \langle f|V|k\rangle\langle k|V|i\rangle \lim_{\varepsilon \to +0}\int_0^\infty dt\left\{e^{i(E_f - E_k + i\varepsilon)t} - e^{i(E_k - E_i + i\varepsilon)t}\right\}. \quad (3.78)$$

Providing the integration over time, we arrive at:

$$\langle f|H^{(2)}|i\rangle = \frac{1}{2}\lim_{\varepsilon \to +0}\int d^3k\, \langle f|V|k\rangle\langle k|V|i\rangle\left(\frac{1}{E_i - E_k + i\varepsilon} + \frac{1}{E_f - E_k + i\varepsilon}\right). \quad (3.79)$$

Now, to get link with the second term in the r.h.s. of Eq. (3.74) it is sufficient to require the energy conservation $E_i = E_f$:

$$\langle f|H^{(2)}|i\rangle\Big|_{E_i = E_f} = \langle f|t^{(2)}|i\rangle, \quad (3.80)$$

where $t^{(2)} = \lim_{\varepsilon \to +0}\int d^3k\, \dfrac{\langle f|V|k\rangle\langle k|V|i\rangle}{E_i - E_k + i\varepsilon}$.

This observation means that the following value

$$\frac{1}{2}\lim_{\varepsilon\to+0}\int d^3k \langle f|V|k\rangle\langle k|V|i\rangle\left(\frac{1}{E_i-E_k+i\varepsilon}-\frac{1}{E_f-E_k+i\varepsilon}\right), \tag{3.81}$$

which distinguishes our perturbation approach from the OFPT in the second order in $V$ conditions the appearance of the second order off-energy-shell structures in the respective operators of physical interactions in a natural way. It is important to emphasize that these structures clearly cannot appear when calculating the $t$-matrix via the Dyson-Feynman method.

In the r.h.s. of Eq. (3.75) the matrix element of the third order in $V$ has the form:

$$\langle f|H^{(3)}|i\rangle = \frac{1}{3}\langle f|RRV-2RVR+VRR|i\rangle$$

$$=-\frac{1}{3}\lim_{\substack{\varepsilon_1\to+0 \\ \varepsilon_2\to+0}}\int_0^\infty dt_1 \int_0^\infty dt_2 e^{-(\varepsilon_1+\varepsilon_2)t}\langle f|V(t_1)V(t_2)V-2V(t_1)VV(t_2)+VV(t_1)V(t_2)|i\rangle. \tag{3.82}$$

Following the same recipe that was used for the calculations of the matrix element $\langle f|H^{(2)}|i\rangle$, we find

$$\langle f|H^{(3)}|i\rangle = \frac{1}{3}\lim_{\substack{\varepsilon_1\to+0 \\ \varepsilon_2\to+0}}\int d^3k_1 d^3k_2 \langle f|V|k_1\rangle\langle k_1|V|k_2\rangle\langle k_2|V|i\rangle$$

$$\times\left[\frac{1}{(E_f-E_{k_1}+i\varepsilon_1)(E_{k_1}-E_{k_2}+i\varepsilon_2)}-\frac{2}{(E_f-E_{k_1}+i\varepsilon_1)(E_{k_2}-E_i+i\varepsilon_2)}\right.$$

$$\left.+\frac{1}{(E_{k_1}-E_{k_2}+i\varepsilon_1)(E_{k_2}-E_i+i\varepsilon_2)}\right]. \tag{3.83}$$

Further, imposing the requirement of energy conservation, we get the link:

$$\langle f|H^{(3)}|i\rangle\Big|_{E_i=E_f} = \langle f|t^{(3)}|i\rangle, \qquad (3.84)$$

where $t^{(3)} = \lim\limits_{\substack{\varepsilon_1\to+0 \\ \varepsilon_2\to+0}} \int d^3k_1 d^3k_2 \dfrac{\langle f|V|k_1\rangle\langle k_1|V|k_2\rangle\langle k_2|V|i\rangle}{\left(E_i - E_{k_1} + i\varepsilon_1\right)\left(E_i - E_{k_2} + i\varepsilon_2\right)}.$

Proceeding this way to an arbitrary order of $V$, we can note the general relation:

$$\langle f|H^{(n)}|i\rangle\Big|_{E_i=E_f} = \langle f|t^{(n)}|i\rangle. \qquad (3.85)$$

The difference $\langle f|H^{(n)}|i\rangle - \langle f|t^{(n)}|i\rangle$ obviously carries out the off energy shell contributions in $\langle f|H^{(n)}|i\rangle$.

We see that the perturbation series that arises in our approach after the first clothing transformation coincides on the energy shell with the series for the *t*-operator in the OFPT. This means that the analytical expressions for the mass and charge shifts and the operators of physical processes derived in the first non-vanishing orders in the coupling constant in the method of unitary clothing transformations will coincide on the energy shell with the respective expressions found within the covariant Dyson-Feynman perturbation theory and no additional detailed proof is necessary.

At the same time, as we have shown, starting form the fourth order in the coupling constant, the actual matrix elements of the *t*-operator derived in our approach will differ from those ones found with help of Feynman rules. In fact, the second and subsequent clothing transformations will bring on additional contributions to the total Hamiltonian, generating new off-energy-shell-structures in the operators of physical processes and providing more consistent consideration of the relativistic effects in the respective scattering amplitudes.

# CHAPTER 4
# DISCUSSION OF THE FOUNDATIONS
# OF THE UNITARY CLOTHING TRANSFORMATION METHOD
# AND THE RESULTS OF ITS APPLICATION

In the present master thesis the unitarily equivalent formulation of the quantum field theory after Greenberg and Schweber [1], in which several essential difficulties typical of conventional relativistic quantum theories of interacting fields are avoided thanks to the reformulation of one and the same total field Hamiltonian operator on new Fock space of clothed (physical) hadronic states, is investigated.

Two fundamental difficulties of traditional formulations of quantum field theory are mirrored in the structure and properties of operators usually dealt with. The first difficulty is connected with the assumption of the local character of interactions between fields and conditions the appearance of divergences both in, e.g., the operators performing mass renormalization and the operators of physical interactions. The usual way to overcome this drawback is the introduction of some Lorenz-scalar functions (phenomenological cut-off form-factors) in each interaction vertex, the existence of which does not violate basic symmetries of the theory (e.g., C, CP, CPT, etc., depending on the type of interaction), and, of cause, provides the convergence of all the integrals determining the observable values. It is clear that there are no fundamental reasons for choosing some particular form of such functions in the framework of, say, quantum mesodynamics. However, one should not expect that the successes in quantum chromodynamics will help solving the problem in future. Apparently, the only way out from this situation could be more deep investigation of any possible attempts to develop the theories of non-local fields and non-local interactions between them. Unfortunately, the problems of obeying the basic symmetries in such theories are still open.

The second difficulty originates from the very structure of the field theory itself even before the procedure of canonical quantization (e.g., after Dirac). Really, all couplings between classical fields, permissible by the special relativity, under quantization, i.e. replacing the classical fields by the operators and there further expansion into plane waves, generate, after the normal ordering, all of the combinatorially possible combinations of the creation/destruction operators. This fact argues, at least intuitively, in favor of Weinberg's statement that any operator in the Hilbert space of states of field quanta can be presented as the infinite sum of the infinite products of creation/destruction operators (of cause, normally ordered), the kernel of which carries information about the translational invariance of the theory, i.e., the $\delta$-function reflecting the 3-momentum conservation in the closed system [26]. Bearing in mind that the $S$-operator is determined as the operator of evolution of the system from the distant past to the distant future, it is no wonder that the majority of operator structures generated in the $S$-operator by the initial interaction Hamiltonian corresponds to the processes which cannot hold in nature. Those are operators responsible for the disconnected graphs and the processes having no energy shell, including the operators generating clouds of virtual quanta [2,3] surrounding free particles, etc.

From the mathematical viewpoint, the infinite number of degrees of freedom conditioned by such a structure of the $S$-operator generates the infinitely dimensional space of all combinatorially possible states of particles. Unfortunately, the application of the well-developed methods of quantum mechanics to the eigenvalue problem for some operator corresponding to some physical value is not grounded. The matter is that the rotation in the space with aid to diagonalyze the matrix of operator can be correctly performed only in the finitely dimensional spaces. In the case the space dimension is infinite or even not determined the only reasonable way to solve the eigenvalue problem is to replace the initial space in which the operator in question has a non-diagonal form by some new orthogonal space in which the operator would have the sparse (block), i.e., approximately diagonal, structure giving at least approximate solution to the problem. Taking into account that the space of states is built by means of the same field quanta as the *S*-operator is, searching for the new space of states appears equivalent to such a choice of the creation/destruction operators in terms of which both the Hamiltonian and the *S*-operator do not contain all the spectrum of operators having no reasonable physical interpretation. This means, that the solution of the eigenvalue problem in quantum field theory can be treated just as a purely algebraic procedure of manipulating with the creation/destruction operators without any connection with any space of states. Therefore, it looks natural to treat the Hamiltonian apart from the space of states and consider it (just as any other operators) as an element of some algebra that is formed by the creation/destruction operators themselves. In this case, addition, subtraction and multiplication are the algebraic operations of the algebra. Besides, this algebra should contain a specific unitary element the structure of which is chosen due to the requirement of absence of certain undesirable elements in the algebra. Thus, the algebra (in the above-mentioned sense) can be juxtaposed to the initial system of interacting fields, and the eigenvalue problem of the Hamiltonian can be investigated as a problem of finding the unitary element of the algebra [9]. The specifics of the unitary element are connected with the fact that it contains all information about those elements eliminated from the algebra. Having cleaned the algebra from those undesirable elements, i.e., fixed the structure of the unitary element, we automatically obtain, say, the Hamiltonian operator (and thus the scattering operator) in the form free from the discussed disadvantages.

What should be the structure of the unitary element of the algebra? To answer the question one needs to have information on the types of elements existing in the algebra from the very beginning. Obviously, the algebra contains elements corresponding to the physical observable processes. Also the algebra contains elements to which the one-particle operators (e.g., the free part of the Hamiltonian) can be juxtaposed. Besides, according to Van Hove [2,3], the algebra comprises the elements generating clouds of virtual field quanta sur-rounding the bare particles. Which of these elements of algebra lead to the problems we are concerned with?

As mentioned above, the divergences in the one-particle operators are tackled with help of the introduction of the cut-off form factors, so that the elements of such a kind need no elimination. Therefore, all attention should be paid to the clouds of virtual quanta studied by Van Hove. We stress, that these multiparticle effects are associated exclusively with the creation/destruction operators themselves but not the structures (e.g., the operators of processes) built out of them. Solely due to that reason Greenberg and Schweber have proposed to seek for the unitary element of the algebra in question in a way the one-particle states in some initially undefined Hilbert space be the eigenstates of the total Hamiltonian. It can be shown (see Sec. 2.5) that the unitary element chosen according to the requirements of the Okubo approach and the Haitler, Sato and Lee method eliminates not only the clouds of virtual quanta but also the part of a subclass of elements associated with the observable processes. Consequently, just the method of the clothing UT appears the most adequate not only from the mathematical but also the physical viewpoint.

According to Greenberg and Schweber, the goal of clothing is the construction of the new Fock space of states in which the clothed (physical) one-particle states appear the eigenstates of the total Hamilonian thanks to the UT of the bare (primary) creation/destruction operators of bare particles. Under this UT, all operator structures of the initial Hamiltonian which prevent the one-particle states to be its eigenstates (namely these operators create clouds of virtual particles) appear accumulated into the new clothed (physical) creation/destruction operators: one can show that the clothed operators are expressed through the infinite sums of infinite products of initial (bare) creation/destruction operators of particles.

The deliverance of the undesirable ("bad" in our treatment) operators is performed via the proper choice of the generators of UT's of the creation/destruction operators, consequently order by order in the coupling constant.

Thus, the procedure has recursive character. This is conditioned by the fact that the explicit form of the operators to be eliminated in some *n*-th order in the coupling constant can be established only after all the bad terms of the lower orders are removed. We emphasize that the elimination of bad operators does not mean their total disappearance from the Hamiltonian. On the contrary, the analyses of the several first clothing UT's witnesses that all eliminated operators appear "built into" the operators of physical processes in higher orders. In other words, the amount of information put into the initial Hamiltonian remains unchanged but "redistributed" according to the physical requirements imposed upon the system of interacting fields under study.

Besides the bad operators, in the course of clothing, the Hamiltonian contains good operators of even orders that repeat the structure of the free part of the Hamiltonian. The kernels of these operators determine the corrections for the self-energies of particles associated with the effects of self-action. In other words, the values of the initial (bare) masses get corrections to their values and finally, if the clothing procedure has been successful, tend to the observable values. The major inconvenience arising with the mass renormalization performed that way is associated with the fact that all observable values are finally expressed through the bare masses that are unknown. This difficulty can be avoided if from the very beginning, before starting the elimination procedure, one performs the auxiliary UT from the creation/destruction operators of bare particles with unphysical masses to the same operators but with physical masses. As the result, the free part of the Hamiltonian gives rise to the new additional operator (mass counterterm), which defines the mass shifts of new particles but no longer conserves their number. Further, in the course of clothing, the arising one-particle operators are cancelled (at least, partly) with the same operators from the mass counterterm forming the mass corrections *post factum*. Finally, this allows expressing all operators of physical processes through the physical masses of particles. Note, that the described auxiliary mass changing UT does not alter the form of the initial interaction operator and appears equivalent to the transition to such new space of states (orthogonal to the initial one) in which the number of free bare particles with physical masses is not conserved.

The problem of coupling constant renormalization fundamentally differs from the similar problem for the mass renormalization. The point is that in any RQFT model the primary interaction operator consists, at least partly, of bad terms. In the model of the Yukawa-type interaction all the vertex operators in the Hamiltonian are bad and, according to the first principles of clothing, they must be removed from the Hamiltonian forming the clothed (renormalized) vertices in physical operators. The other approach, which is analogous, in part, to that one used in the mass renormalization, is also possible. Similar to the introduction of the mass counterterm, one can introduce the vertex counterterm defining the charge correction. With help of such a trick, in this work we have performed the transition to the observable value of the coupling constant while the respective correction is found *post factum* in the course of clothing. The weak point in this scheme is the fact that the determination of charge correction is performed via the cancellation of bad operators. One of them originates from the vertex counterterm (which is bad by definition) while the other arises during the clothing and stems from the multiple commutators, and is bad too. This trouble manifests itself in the way in this work we perform the comparison with explicitly covariant calculations of the correction in question in the third order in the coupling constant. Really, the matrix element of the $S$-operator between one- and two-particle states is equal to zero by definition because the mentioned transition is kinematically forbidden. Therefore, instead of performing direct but ambiguous calculations we establish some general links between the perturbation series appearing in the unitary clothing transformation approach and the series typical of the OFPT which is in fact equivalent to the purely covariant time-dependent Dyson-Feynman perturbation theory.

However, the mentioned circumstance that seems a drawback for the models where the interaction operator is totally bad could appear an advantage for the model where the interaction operator also contains good part. Then the renormalization of the coupling constant which determines the bad part of the interaction operator can be realized as the vertex clothing in the operators of physical processes while the shift of charge which is associated with the good part the interaction operator can be determined with help of introduction and subsequent cancellation of the vertex counterterm.

In general, the charge shift is usually defined by comparing two certain values generated by two mechanisms of one and the same physical process (say, πN→πN). One of these mechanisms contains one interaction vertex renormalized to the certain order in the physical coupling constant. In the covariant Dyson-Feynman formalism those values are the scattering amplitudes. They are defined on the energy shell, so the difference between them, which determines the charge shift, is also a value on the energy shell. Contrary to the covariant approach, in the method of clothing the role of the mentioned values are played by the operators of physical interactions in the Hamiltonian, which are determined off the energy shell. Consequently, the charge shift found by clothing is the object off the energy shell. For example, if one calculates the expression for charge correction via comparing two operators of physical process πN→πN in the second and fourth orders, one finds this expression equivalent to that one obtained in this work via the cancellation of the vertex-like operators. If from the very beginning these operators are projected on the energy shell, then the expression for the charge shift coincides with that one derived on the energy shell. That is why we divide the expression for $\delta g^{(3)}$ into two parts: $\delta g^{(3)}_{off-energy-shell}$, which goes to zero on the energy shell, and $\delta g^{(3)}_{on-energy-shell}$, which can be brought to the explicitly covariant form, providing the independence of the charge shift of the particle momenta.

Finally, if the clothing procedure has been successfully completed, i.e., the Hamiltonian and all other generators of the Poincare group are free from bad operators and the mass and charge renormalization programs are realized, then the operators remaining in the Hamiltonian present only observable (physical) processes. Such operators are naturally Hermitian and relativistic, independent of the interaction energy, include the recoil effects (and non-local in this sense), contain the off-energy-shell structures and are defined in the Hilbert space of the physical many-particle states. Further, to calculate the observables, these operators should be put into the relativistic dynamical equations of the Lippmann-Schwinger type for the $t$-matrix, which are solved via the usual methods of the theory of integral equations. During the solution of these equations, the summation of all the infinite series of mechanisms responsible for the investigated process is performed. The $t$-matrix derived in such way defines the respective $S$-matrix and, finally, the amplitude of the physical process studied. It is remarkable that from the moment we have obtained the operator of some physical process with help of clothing, all the remaining steps towards the derivation of observables can be performed using the well-developed methods of nuclear physics.

In conclusion we note that the present unitarily equivalent formulation of the relativistic quantum field theory can be considered as the result of application of the specific noncovariant off-energy-shell perturbation theory in which all of observable results are determined via the operators only on the 3-momentum shell. Due to this feature of our approach, there arises a possibility of consistent consideration of the relativistic effects off the energy shell. The latter are expected to bring on essential contributions to the amplitudes of observable processes even in case of low energies, and also can help to investigate the properties of not only the bound states of a few-nucleon system but of the nuclear matter as well.